\documentstyle[stwol,psfig]{article}

\newcommand{\epe}{\varepsilon'/\varepsilon}
\newcommand{\mt}{m_{\rm t}}
\newcommand{\mtb}{\overline{m}_{\rm t}}
\newcommand{\mc}{m_{\rm c}}
\newcommand{\ms}{m_{\rm s}}

\newcommand{\mb}{m_{\rm b}}
\newcommand{\mw}{M_{\rm W}}
\newcommand{\mz}{M_{\rm Z}}
\newcommand{\gev}{\, {\rm GeV}}
\newcommand{\mev}{\, {\rm MeV}}

\newcommand{\Lms}{\Lambda_{\overline{\rm MS}}}

\newcommand{\Bsg}{$B \to X_s \gamma$ }

\newcommand{\RE}{{\rm Re}}
\newcommand{\IM}{{\rm Im}}

\newcommand{\vcb}{|V_{cb}|}
\newcommand{\vtd}{|V_{td}|}
\newcommand{\vub}{|V_{ub}/V_{cb}|}
\newcommand{\vts}{|V_{ts}|}

\newcommand{\bea}{\begin{eqnarray}}
\newcommand{\eea}{\end{eqnarray}}
\newcommand{\bd}{\begin{displaymath}}
\newcommand{\ed}{\end{displaymath}}

\begin{document}

\twocolumn[
\begin{flushright}
TUM-HEP-259/96 \\
MPI-PhT/96-111 \\
hep-ph/9610461 \\
October 1996
\end{flushright}

\vskip1truecm

\centerline{\bf FLAVOUR CHANGING NEUTRAL CURRENT PROCESSES${}^\dagger$}

\vskip1truecm

\centerline{\bf Andrzej J. Buras}
\bigskip
\centerline{\sl Technische Universit\"at M\"unchen, Physik Department}
\centerline{\sl D-85748 Garching, Germany}
\vskip0.6truecm
\centerline{\sl Max-Planck-Institut f\"ur Physik}
\centerline{\sl  -- Werner-Heisenberg-Institut --}
\centerline{\sl F\"ohringer Ring 6, D-80805 M\"unchen, Germany}

\vskip1truecm
\thispagestyle{empty}

\centerline{\bf Abstract}

We review the status of the flavour-changing neutral-current 
processes (FCNC). In particular we discuss:
i) Main targets of the field,
ii) The theoretical framework for FCNC, 
iii) Standard analysis of the unitarity triangle, 
iv) $\varepsilon'/\varepsilon$,
v) Radiative, rare and CP-violating decays, 
vi) CP-asymmetries in B-decays,
vii) Comparision of the potentials of
$ K \to \pi\nu\bar\nu $ and CP-B asymmetries,
vii) Some aspects of the physics beyond the Standard Model.

\vspace{6cm}

${}^\dagger${\footnotesize 
\noindent Plenary Talk given at
             the 28th International Conference on High Energy Physics,
             Warsaw, Poland, 25-31 July 1996, to appear in the proceedings.\\
             Supported by the German
             Bundesministerium f\"ur Bildung and Forschung under contract
             06 TM 743 and DFG Project Li 519/2-1.}
]


\setcounter{page}{0}

\title{FLAVOUR CHANGING NEUTRAL CURRENT PROCESSES}

\author{Andrzej J. Buras}

\address{
Technische Universit\"at M\"unchen, Physik Department \\
D-85748 Garching, Germany \\
\vskip 0.3cm
Max-Planck-Institut f\"ur Physik \\
-- Werner-Heisenberg-Institut -- \\
F\"ohringer Ring 6, D-80805 M\"unchen, Germany
}

\twocolumn[\maketitle\abstracts{
We review the status of the flavour-changing neutral-current 
processes (FCNC). In particular we discuss:
i) Main targets of the field,
ii) The theoretical framework for FCNC, 
iii) Standard analysis of the unitarity triangle, 
iv) $\varepsilon'/\varepsilon$,
v) Radiative, rare and CP-violating decays, 
vi) CP-asymmetries in B-decays,
vii) Comparision of the potentials of
$ K \to \pi\nu\bar\nu $ and CP-B asymmetries,
vii) Some aspects of the physics beyond the Standard Model.
}]


\section{Introduction}
The flavour-changing neutral-current processes (FCNC) such as 
particle-antiparticle mixing, certain rare 
and radiative meson decays and 
CP-violating decays have played an important role in the construction
of the standard model. In this model they are governed by the
Glashow-Iliopoulos-Maiani (GIM) mechanism \cite{GIM} which assures a
natural suppression of these processes, the fact observed experimentally.
As a consequence of this mechanism, 
there are no FCNC processes at the tree
level and the leading contributions result from the one-loop diagrams:
the penguin and the box diagrams (fig. 1). 
The latter fact makes FCNC processes
a very efficient tool for the determination of certain parameters of
the Cabibbo-Kobayashi-Maskawa matrix \cite{CAB,KM}, in particular the
top-quark couplings $|V_{td}|$ and $|V_{ts}|$ and the CP-violating
phases. Simultaneously FCNC are sensitive to contributions from physics
beyond the standard model.

At present there are only a few FCNC transitions which have been observed
experimentally. These are:
\begin{itemize}
\item
$K^0-\bar K^0$ mixing, the related $K_L-K_S$ mass difference and the
indirect CP violation in $K_L\to \pi\pi$ represented by the parameter
$\varepsilon_K$.
\item
$B_d^0-\bar B_d^0$ mixing and the related mass difference $(\Delta M)_d$.
\item
$K_L \to \mu\bar\mu$.
\item
$B\to X_s\gamma$ and $B\to K^*\gamma$.
\end{itemize}
There are also controversial experimental results on the direct CP
violation in $K_L\to \pi\pi$ represented by the ratio $\epe$ to which
we will return below. For the remaining FCNC transitions only upper
bounds exist which in many cases are still several orders of magnitude
above the standard model expectations. Moreover the very spare experimental 
information on CP violation leaves the issue of this important phenomenon
completely open. This situation should change considerably during
the next ten years due to forthcoming experiments at $e^+e^-$ B-factories,
HERA-B, dedicated K-physics experiments at CERN, BNL, KEK and DA${\Phi}$NE,
and efforts at hadron colliders such as TEVATRON and LHC.

\psscalefirst

\begin{figure}
\centerline{
\psfig{figure=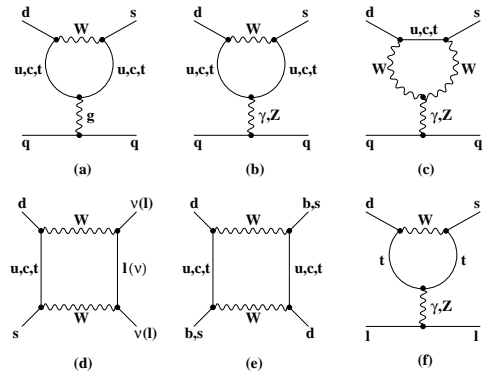,height=2.5in,angle=270}}
\caption[]{Typical Penguin and Box Diagrams.
\label{fig:fdia}}
\end{figure}

The purpose of this review is to summarize the present status of the FCNC
processes and to provide an outlook for future developments.

Section 2 gives a " Grand View" of the field, discussing 
its most important targets, recalling the CKM matrix and the unitarity
triangle and presenting briefly the theoretical framework.
Section 3 discusses the by now standard analysis of the unitarity triangle
(UT). 
Section 4 summarizes the present status of  $\epe$.
Section 5 summarizes the present status of radiative and rare B-decays. 
Section 6  summarizes the present status of rare K-decays.
Section 7 summarizes the present status of CP asymmetries in B-decays.
Section 8 begins with a classification of  K- and B-decays from the point
of view of theoretical cleanliness.
Subsequently  the potentials of CP asymmetries in B-decays and of
the very clean decays $K^+\to\pi^+\nu\bar\nu$ and $K_L\to\pi^o\nu\bar\nu$ 
in determining the parameters of the CKM matrix are compared.
Section 9 offers a brief look beyond the Standard Model.
Section 10  gives a short summary and an outlook.
Except for sections 5,7 and 10 there is a considerable overlap
between this review and another review \cite{AJB96}.  

\section{Grand View}
\subsection{Main Targets of FCNC Processes}
These are:
\begin{itemize}
\item
The parameters of the CKM matrix. In particular: the parameters 
$\eta$, $\varrho$ and the elements $|V_{td}|$  and $|V_{ts}|$,
\item
CP violation in the Standard Model,
\item
Physics beyond the Standard Model such as supersymmetry, left-right
symmetry, charged higgs scalars, leptoquarks, lepton number violations
etc.,
\end{itemize}
\subsection{The CKM Matrix and the Unitarity Triangle}
An important target of particle physics is the determination
 of the unitary $3\times 3$ Cabibbo-Kobayashi-Maskawa
matrix \cite{CAB,KM} which parametrizes the charged current interactions of
 quarks:
\begin{equation}\label{1j}
J^{cc}_{\mu}=(\bar u,\bar c,\bar t)_L\gamma_{\mu}
\left(\begin{array}{ccc}
V_{ud}&V_{us}&V_{ub}\\
V_{cd}&V_{cs}&V_{cb}\\
V_{td}&V_{ts}&V_{tb}
\end{array}\right)
\left(\begin{array}{c}
d \\ s \\ b
\end{array}\right)_L
\end{equation}
The CP violation in the standard model is supposed to arise
from a single phase in this matrix.
It is customary these days to express the CKM-matrix in
terms of four Wolfenstein parameters 
\cite{WO} $(\lambda,A,\varrho,\eta)$
with $\lambda=\mid V_{us}\mid=0.22 $ playing the role of an expansion 
parameter and $\eta$
representing the CP violating phase:
\begin{eqnarray}
\lefteqn{V_{CKM} =}
\nonumber \\
& &
\left(\begin{array}{ccc}
1-{\lambda^2\over 2}&\lambda&A\lambda^3(\varrho-i\eta)\\ -\lambda&
1-{\lambda^2\over 2}&A\lambda^2\\ A\lambda^3(1-\varrho-i\eta)&-A\lambda^2&
1\end{array}\right)
\nonumber \\
& & +~O(\lambda^4)
\label{2.75} 
\end{eqnarray}
Because of the
smallness of $\lambda$ and the fact that for each element 
the expansion parameter is actually
$\lambda^2$, it is sufficient to keep only the first few terms
in this expansion. 

Following \cite{BLO} one can define the parameters
$(\lambda, A, \varrho, \eta)$ through
\begin{equation}\label{wop}
s_{12}\equiv\lambda \quad s_{23}\equiv A \lambda^2 \quad
s_{13} e^{-i\delta}\equiv A \lambda^3 (\varrho-i \eta)
\end{equation}
where $s_{ij}$ and $\delta$ enter the standard exact 
parametrization \cite{PDG}  of the CKM
matrix. This specifies the higher orders terms in (\ref{2.75}).

The definition of $(\lambda,A,\varrho,\eta)$ given in (\ref{wop})
is useful because it allows to improve the accuracy of the
original Wolfenstein parametrization in an elegant manner. In
particular
\begin{equation}\label{CKM1}
V_{us}=\lambda \qquad V_{cb}=A\lambda^2
\end{equation}
\begin{equation}\label{CKM2}
V_{ub}=A\lambda^3(\varrho-i\eta)
\qquad
V_{td}=A\lambda^3(1-\bar\varrho-i\bar\eta)
\end{equation}
where
\begin{equation}\label{3}
\bar\varrho=\varrho (1-\frac{\lambda^2}{2})
\qquad
\bar\eta=\eta (1-\frac{\lambda^2}{2})
\end{equation}
turn out \cite{BLO} to be excellent approximations to the
exact expressions. 

\begin{figure}
\centerline{
\psfig{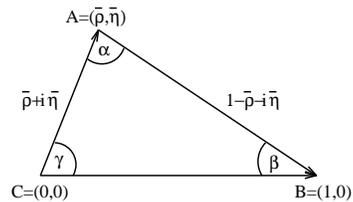}}
\caption[]{Unitarity Triangle.
\label{fig:utriangle}}
\end{figure}

A useful geometrical representation of the CKM matrix is the unitarity 
triangle obtained by using the unitarity relation
\begin{equation}\label{2.87h}
V_{ud}V_{ub}^* + V_{cd}V_{cb}^* + V_{td}V_{tb}^* =0,
\end{equation}
rescaling it by $\mid V_{cd}V_{cb}^\ast\mid=A \lambda^3$ and depicting
the result in the complex $(\bar\rho,\bar\eta)$ plane as shown
in fig. 2. The lenghts CB, CA and BA are equal respectively to 1,
\begin{eqnarray}
R_b &\equiv&  \sqrt{\bar\varrho^2 +\bar\eta^2}
= (1-\frac{\lambda^2}{2})\frac{1}{\lambda}
\left| \frac{V_{ub}}{V_{cb}} \right|
\nonumber \\
R_t &\equiv& \sqrt{(1-\bar\varrho)^2 +\bar\eta^2}
=\frac{1}{\lambda} \left| \frac{V_{td}}{V_{cb}} \right|.
\label{2.94a}
\end{eqnarray}

The triangle in fig. 2, $\mid V_{us}\mid$ and $\mid V_{cb}\mid$
give the full description of the CKM matrix. 
Looking at the expressions for $R_b$ and $R_t$ we observe that within
the standard model the measurements of four CP
{\it conserving } decays sensitive to $|V_{us}|$, $\vcb$,   
$| V_{ub}| $ and $ |V_{td}|$ can tell us whether CP violation
($\eta \not= 0$) is predicted in the standard model. 
This is a very remarkable property of
the Kobayashi-Maskawa picture of CP violation: quark mixing and CP violation
are closely related to each other. 

There is of course the very important question whether the KM picture
of CP violation is correct and more generally whether the standard
model offers a correct description of weak decays of hadrons. In order
to answer these important questions it is essential to calculate as
many branching ratios as possible, measure them experimentally and
check if they all can be described by the same set of the parameters
$(\lambda,A,\varrho,\eta)$. In the language of the unitarity triangle
this means that the various curves in the $(\bar\varrho,\bar\eta)$ plane
extracted from different decays should cross each other at a single point
as shown in fig. 3.
Moreover the angles $(\alpha,\beta,\gamma)$ in the
resulting triangle should agree with those extracted one day from
CP-asymmetries in B-decays.  

\begin{figure}
\centerline{
\psfig{figure=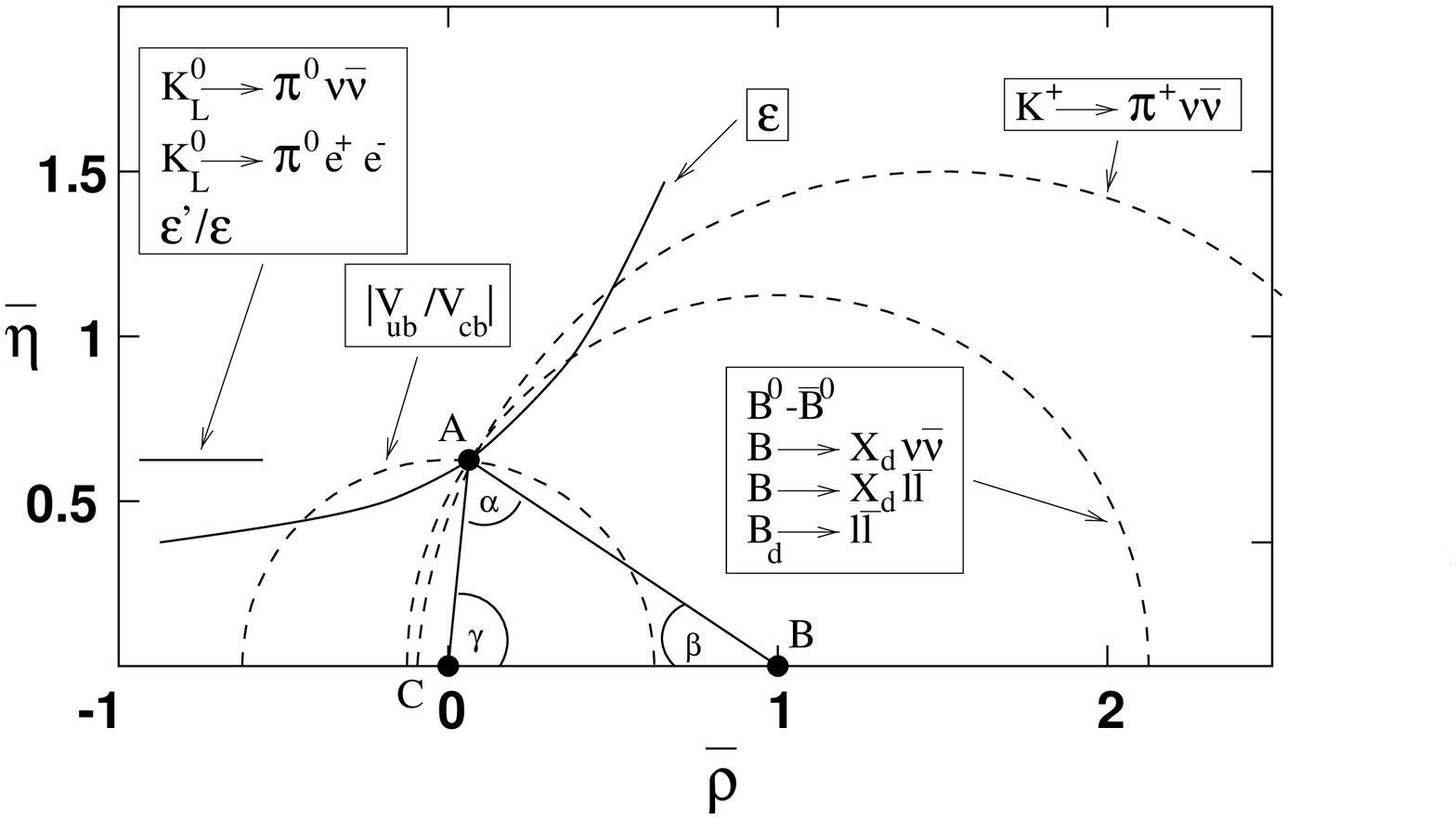,height=3.5in,angle=270}}
\caption[]{The ideal Unitarity Triangle. For artistic reasons the value of
$\bar\eta$ has been chosen to be higher than the fitted central value
$\bar\eta\approx 0.35.$
\label{fig:2011}}
\end{figure}

Since the CKM matrix is only a parametrization of quark mixing and 
of CP violation and does not offer the explanation of these two
very important phenomena, many physicists hope that a new physics
while providing a dynamical origin of quark mixing and of 
CP violation will
also change the picture given in fig. 3. That is, the different curves
based on standard model expressions, will not cross each other 
at a single point
and the angles $(\alpha,\beta,\gamma)$ 
extracted one day from
CP-asymmetries in B-decays will disagree with the ones determined from
rare K and B decays.

Clearly the plot in fig. 3 is highly idealized because in order
to extract such nice curves from various decays one needs perfect
experiments and perfect theory. We will see below that 
for certain decays such a picture is not fully unrealistic.
Generally however the task of extracting the unitarity triangle
from the experimental data is difficult.
Here are the reasons.

\subsection{Theoretical Framework}
The basic problem in the calculation of branching ratios for hadron
 decays
and other physical observables is related to strong 
interactions. Although due to  
the smallness
of the effective QCD coupling at short distances, the gluonic
contributions at scales ${\cal O} (\mw, \mz, \mt)$ can be calculated
within the perturbative framework, the fact that
mesons are $ q\bar q$ bound states forces us to consider  QCD at 
long distances as well.
 Here we have to rely on existing non-perturbative
methods which are not yet very powerful
at present.

The Operator Product Expansion (OPE) combined with the renormalization group
approach  allows to divide the problem
into two parts: the short distance part, under control already today,
and the long distance part which hopefully will be fully under control
when our non-perturbative tools improve.
This framework 
brings in local operators $Q_i$ which govern ``effectively''
the transitions in question and 
the amplitude for an {\it exclusive} decay $M\to F$ 
is written as
\begin{equation}\label{OPE}
 A(M \to F) = \frac{G_F}{\sqrt 2} {\rm V_{CKM}} \sum_i
   C_i (\mu) \langle F \mid Q_i (\mu) \mid M \rangle 
\end{equation}
where $M$ stands for the decaying meson, $F$ for a given final state and
${\rm V_{CKM}}$ denotes the relevant CKM factor.
$ Q_i(\mu) $ denote
the local operators generated by QCD and electroweak interactions.
$ C_i(\mu) $ stand for the Wilson
coefficient functions (c-numbers). 
The scale $ \mu $ separates the physics contributions in the ``short
distance'' contributions (corresponding to scales higher than $\mu $)
contained in $ C_i(\mu) $ and the ``long distance'' contributions
(scales lower than $ \mu $) contained in 
$\langle F \mid Q_i (\mu) \mid M \rangle $.
The $ (\mu -dependence) $ of $ C_i(\mu) $ is governed by
the renormalization group equations.
This $ \mu $-dependence 
must be cancelled by the one present in $\langle  Q_i (\mu)\rangle $
so that the full amplitude does not depend on $ \mu $.
Generally this cancellation involves many
operators due to the operator mixing under renormalization.

It should  be stressed that the use of the renormalization group
is  necessary in order to sum up large logarithms
 $ \log \mw/\mu $ which appear for $ \mu= {\cal O}(1-5\gev) $.
 In the so-called leading
logarithmic approximation (LO) terms $ (\alpha_s\log \mw/\mu)^n $ are summed.
The next-to-leading logarithmic correction (NLO) to this result involves
summation of terms $ (\alpha_s)^n (\log \mw/\mu)^{n-1} $ and so on.
This hierarchic structure gives the renormalization group improved
perturbation theory.

I will not discuss here the technical details of the renormalization group
and of the calculation of
$C_i(\mu)$. They can be found in a recent review \cite{BBL}.
Let me just list a few operators which play an important role
in the phenomenology of weak decays. These are ($\alpha$ and $\beta$ are
colour indices):

\noindent
{\bf Current--Current:}
\begin{eqnarray}
Q_1 &=& (\bar s_{\alpha} u_{\beta})_{V-A}\;(\bar u_{\beta} d_{\alpha})_{V-A}
\nonumber \\
Q_2 &=& (\bar s u)_{V-A}\;(\bar u d)_{V-A} 
\label{O1} 
\end{eqnarray}

\noindent
{\bf QCD--Penguin and Electroweak--Penguin:}
\begin{eqnarray}
Q_6 = (\bar s_{\alpha} d_{\beta})_{V-A}\sum_{q=u,d,s}
       (\bar q_{\beta} q_{\alpha})_{V+A}  &&
\nonumber \\
Q_8 = {3\over2}\;(\bar s_{\alpha} d_{\beta})_{V-A}\sum_{q=u,d,s}e_q
        (\bar q_{\beta} q_{\alpha})_{V+A} &&
\label{O2}
\end{eqnarray}

\noindent
{\bf Magnetic--Penguins:}
\begin{eqnarray}
Q_{7\gamma}  =  \frac{e}{8\pi^2} m_b \bar{s}_\alpha \sigma^{\mu\nu}
          (1+\gamma_5) b_\alpha F_{\mu\nu}
&&
\nonumber \\
Q_{8G}     =  \frac{g}{8\pi^2} m_b \bar{s}_\alpha \sigma^{\mu\nu}
   (1+\gamma_5)T^a_{\alpha\beta} b_\beta G^a_{\mu\nu}  
&&
\label{O6}
\end{eqnarray}

\noindent
{\bf $\Delta S = 2 $ and $ \Delta B = 2 $ Operators:}
\begin{eqnarray}
Q(\Delta S = 2)  &=& (\bar s d)_{V-A} (\bar s d)_{V-A}
\nonumber \\
Q(\Delta B = 2)  &=& (\bar b d)_{V-A} (\bar b d)_{V-A} 
\label{O7}
\end{eqnarray}

\noindent
{\bf Semi--Leptonic Operators:}
\begin{equation}\label{9V}
Q_{9V}  = (\bar s d  )_{V-A} (\bar e e)_{V} \
Q_{10A}  = (\bar s d )_{V-A} (\bar e e)_{A}
\end{equation}
\begin{equation}\label{9NU}
Q_{\bar\nu\nu}  = (\bar s d )_{V-A} (\bar\nu\nu)_{V-A} \
Q_{\bar\mu\mu}  = (\bar s d )_{V-A} (\bar\mu\mu)_{V-A}
\end{equation}

The formal expression for the decay amplitudes given in
(\ref{OPE}) can be cast in the form \cite{PBE}:
\begin{equation}\label{PBEE}
A(M\to F)=\sum_i B_i {\rm V_{CKM}}^{i} \eta^{i}_{QCD} F_i(\mt,\mc)
\end{equation}
which is more useful for phenomenology. In writing (\ref{PBEE})
we have generalized (\ref{OPE}) to include several CKM factors.
$F_i(m_t,m_c)$, the Inami-Lim functions \cite{IL}, 
 result from the evaluation of loop diagrams with
internal top and charm exchanges (see fig. 1) and may also depend
solely on $\mt$ or $\mc$. In the case of new physics they  
depend  on masses of new particles such as charginos, stops, charged
Higgs scalars etc. 
The factors $\eta^{i}_{QCD}$ summarize
the QCD corrections which can be calculated by formal methods
discussed above. Finally $B_i$ stand for nonperturbative factors
related to the hadronic matrix elements of the contributing
operators: the main theoretical uncertainty in the whole enterprise.
In leptonic and semi-leptonic decays for which only the matrix elements
of weak currents are needed,  
the non-perturbative $B$-factors can fortunately be determined from
leading tree level decays reducing
or removing the non-perturbative uncertainty. In non-leptonic
decays this is generally not possible and we have to rely on
existing non-perturbative methods. A well known example of a
$B_i$-factor is the renormalization group invariant parameter 
$B_K$  defined by 
\begin{eqnarray}
B_K =B_K(\mu)\left[\alpha_s(\mu)\right]^{-2/9}
 & &  
\label{bk} \\
\langle \bar K^{0}\mid Q(\Delta S=2)\mid K^{0}\rangle &=&
\frac{8}{3} B_K(\mu)F_K^2 m_K^2
\nonumber
\end{eqnarray}
where we did not show the NLO correction in $B_K$.
$B_K$ plays an important role in the phenomenology of CP violation
in $K \to \pi\pi$. 

So far we have discussed only {\it exclusive} decays. During the
recent years considerable progress has been made for inclusive
decays of heavy mesons. The starting point is an effective
hamiltonian with a structure analogous to (\ref{OPE}) in which the short 
distance QCD effects are collected in $C_i(\mu)$. 
The actual decay described by the operators
$Q_i$ is then calculated in the spectator model corrected for
additional virtual and real gluon corrections.
Support for this approximation 
comes from heavy quark ( $1/m_b $) expansions (HQE).
Indeed the spectator
model corresponds to the leading order approximation
in the $1/m_b$ expansion. 
The next corrections appear at the ${\cal O}(1/m_b^2)$
level. The latter terms have been studied by several authors
\cite{Chay,Bj,Bigi} with the result that they affect various
branching ratios by less than $10\%$ and often by only a few percent.

There is a vast literature on this subject and I can only refer here to
a few papers \cite{Bigi,Mannel} where further references can be found.
Of particular importance for this field was also the issue of the
renormalons which are nicely discussed in \cite{Beneke,Braun}.

In order to achieve sufficient precision the Wilson coefficients
or equvalently the QCD factors $\eta^{i}_{QCD}\equiv\eta_i$ have 
to include both
the leading and the next-to-leading (NLO) corrections.
These corrections are known by now
for the most important and interesting decays and are 
reviewed in \cite{BBL}. The list of existing NLO calculations
is given in table 1. 
We will discuss their impact below. 

\begin{table}
\begin{center}
\begin{tabular}{|l|l|}
\hline
\bf \phantom{XXXXXXXX} Decay & \bf Reference \\
\hline
\hline
\multicolumn{2}{|c|}{$\Delta F=1$ Decays} \\
\hline
current-current operators     & \cite{ALTA,BW} \\
QCD penguin operators         & \cite{BJLW1,BJLW,ROMA1,ROMA2} \\
electroweak penguin operators & \cite{BJLW2,BJLW,ROMA1,ROMA2} \\
magnetic penguin operators    & \cite{MisMu:94}  \\
$Br(B)_{SL}$                  & \cite{ALTA,Buch:93,Bagan} \\
inclusive $\Delta S=1$ decays       & \cite{JP} \\
\hline
\multicolumn{2}{|c|}{Particle-Antiparticle Mixing} \\
\hline
$\eta_1$                   & \cite{HNa} \\
$\eta_2,~\eta_B$           & \cite{BJW} \\
$\eta_3$                   & \cite{HNb} \\
\hline
\multicolumn{2}{|c|}{Rare K- and B-Meson Decays} \\
\hline
$K^0_L \rightarrow \pi^0\nu\bar{\nu}$, $B \rightarrow l^+l^-$,
$B \rightarrow X_{\rm s}\nu\bar{\nu}$ & \cite{BB1,BB2} \\
$K^+   \rightarrow \pi^+\nu\bar{\nu}$, $K_L \rightarrow \mu^+\mu^-$
                                      & \cite{BB3} \\
$K^+\to\pi^+\mu\bar\mu$               & \cite{BB5} \\
$K_L \rightarrow \pi^0e^+e^-$         & \cite{BLMM} \\
$B\rightarrow X_s \mu^+\mu^-$           & \cite{Mis:94,BuMu:94} \\
$B\rightarrow X_s \gamma$      & \cite{AG2,Yao1,Pott,GREUB,CZMM} \\
\hline
\end{tabular}
\end{center}
\caption{References to NLO Calculations}
\label{TAB1}
\end{table}

Let us recall why NLO calculations are important for the
phenomenology of weak decays:

\begin{itemize}
\item The NLO is first of all necessary to test the validity of
the renormalization group improved perturbation theory.
\item Without going to NLO the QCD scale $\Lambda_{\overline{MS}}$
extracted from various high energy processes cannot be used 
meaningfully in weak decays.
\item Due to the renormalization group invariance the physical
amplitudes do not depend on the scales $\mu$ present in $\alpha_s$
or in the running quark masses, in particular $\mt(\mu_t)$, 
$\mb(\mu_b)$ and $\mc(\mu_c)$. However
in perturbation theory this property is broken through the truncation
of the perturbative series. Consequently one finds sizable scale
ambiguities in the leading order, which can be reduced considerably
by going to NLO.
\item In several cases the issue of the top quark mass dependence
is strictly a NLO effect.
\end{itemize}

Clearly in order to calculate the full amplitude in (\ref{PBEE}) or 
(\ref{OPE}) also the $B_i$ factors or
the matrix elements $\langle F \mid Q_i (\mu) \mid M \rangle$ have to
be evaluated. Since they involve long distance contributions one is
forced in this case to use non-perturbative methods
such as lattice calculations, the
$1/N$ expansion, QCD sum rules or chiral perturbation theory. In the
case of semi-leptonic B meson decays also the Heavy Quark Effective Theory
(HQET) \cite{NEU} turns out to be a useful tool. 
In HQET the matrix elements are
evaluated approximately in an expansion in $1/m_b$. 
Potential uncertainties in the calculation of the non-leading terms
in this expansion have been stressed recently \cite{MASA}.
Needless 
to say all these non-perturbative methods have some limitations.
 Consequently the dominant theoretical
uncertainties in the decay amplitudes reside in
 $\langle Q_i \rangle$ or the corresponding $B_i$ factors.

\section{Standard Analysis}
\subsection{Basic Formulae}
The standard analysis using the available
experimental and theoretical information proceeds essentially in five
steps:

\noindent
{\bf Step 1:}

\noindent
{}From  $b\to c$ transition in inclusive and exclusive B meson decays
one finds $\vcb$ and consequently the scale of UT:
\begin{equation}
\vcb \Longrightarrow \lambda \vcb= \lambda^3 A
\end{equation}

\noindent
{\bf Step 2:}

\noindent
{}From  $b\to u$ transition in inclusive and exclusive B meson decays
one finds $\vub$ and consequently the side $CA=R_b$ of UT:
\begin{equation}
\vub \Longrightarrow R_b=4.44 \cdot \left| \frac{V_{ub}}{V_{cb}} \right|
\end{equation}

\noindent
{\bf Step 3:}

\noindent
{}From the observed indirect CP violation in $K \to \pi\pi$ described
experimentally by the parameter $\varepsilon_K$ and theoretically
by the imaginary part of the relevant box diagram in fig. 1 one 
derives the constraint:
\begin{eqnarray}
\bar\eta \left[(1-\bar\varrho) A^2 \eta_2 S(x_t)
+ P_0(\varepsilon) \right] A^2 B_K &=& 0.226
\nonumber \\
S(x_t)=0.784 \cdot x_t^{0.76} &&
\label{100}
\end{eqnarray}
where
\begin{equation}\label{102}
P_0(\varepsilon) = 
\left[ \eta_3 S(x_c,x_t) - \eta_1 x_c \right] \frac{1}{\lambda^4}
\quad
x_t=\frac{\mt^2}{\mw^2}
\end{equation}
Equation (\ref{100}) specifies 
a hyperbola in the $(\bar \varrho, \bar\eta)$
plane. Here $B_K$
is the non-perturbative parameter defined in (\ref{bk}) and $\eta_2$
is the QCD factor in the box diagrams with two top quark exchanges.
Finally $P_0(\varepsilon)=0.31\pm0.02$ summarizes the contributions
of box diagrams with two charm quark exchanges and the mixed 
charm-top exchanges. 
The NLO values of the QCD factors $\eta_1$ , $\eta_2$ and $\eta_3$ 
are given as follows \cite{HNa,BJW,HNb}:
\begin{equation}
\eta_1=1.38\pm 0.20 \
\eta_2=0.57\pm 0.01 \
\eta_3=0.47\pm0.04
\end{equation}

The quoted errors reflect the remaining theoretical uncertainties due to
$\Lambda_{\overline{MS}}$ and the quark
masses. 
The factor $\eta_1$ plays only a minor role in the analysis of
$\varepsilon_K$ but its enhanced value through NLO corrections \cite{HNa}
is essential for the $K_L-K_S$ mass difference.

Concerning the parameter $B_K$, the most recent analyses
using the lattice methods summarized by 
Flynn \cite{Flynn} give
$B_K=0.90\pm 0.06$. The $1/N$ approach
 of \cite{BBG0}  gives  $B_K=0.70\pm 0.10$. A recent confirmation of this
result in a somewhat modified framework has been presented by 
Bijnens and Prades \cite{Bijnens}  who gave plausible arguments for 
the difference between this result
for $B_K$ and the lower values obtained by using the QCD Hadronic Duality
approach \cite{Prades} ($B_K=0.39\pm 0.10$) or using the $SU(3)$ symmetry and
 PCAC
($B_K=1/3$) \cite{Donoghue}. For $\vcb=0.040$ and $\vub=0.08$ such 
low values for
$B_K$ require $\mt>200~GeV$ in order to explain the experimental
value of $\varepsilon_K$ \cite{AB,BLO,HNb}. The QCD sum rule results are in
the ball park of $B_K=0.60$ \cite{DENA}.
In the chiral quark model one finds $B_K=0.87\pm0.25$ \cite{BERNEW}.
 In our numerical analysis presented 
below we will use $B_K=0.75\pm 0.15$.

\noindent
{\bf Step 4:}

\noindent
{}From the observed $B^0_d-\bar B^0_d$ mixing described experimentally 
by the mass difference $(\Delta M)_d$ or by the
mixing parameter $x_d=\Delta M/\Gamma_B$ 
and theoretically by the relevant box diagram of fig. 1
the side $BA=R_t$ of the UT can be determined:
\begin{equation}\label{106}
 R_t = 1.0 \cdot
\left[\frac{|V_{td}|}{8.7\cdot 10^{-3}} \right] 
\left[ \frac{0.040}{\vcb} \right]
\end{equation}
with
\begin{eqnarray}
\vtd &=&
8.7\cdot 10^{-3}\left[ 
\frac{200\mev}{\sqrt{B_{B_d}}F_{B_d}}\right]
\left[\frac{170~GeV}{\mtb(\mt)} \right]^{0.76} 
\nonumber \\
& & \times
\left[\frac{(\Delta M)_d}{0.45/ps} \right ]^{0.5} 
\sqrt{\frac{0.55}{\eta_B}}  
\label{VT}
\end{eqnarray}

Here $\eta_B$ is the QCD factor analogous to $\eta_2$ and given
by $\eta_B=0.55\pm0.01$ \cite{BJW}. Next 
$F_{B_d}$ is the B-meson decay constant and $B_{B_d}$ 
denotes a non-perturbative
parameter analogous to $B_K$. 
 
There is a vast literature on the lattice calculations of $F_{B_d}$ 
and $B_{B_d}$.
The most recent world averages are given by Flynn \cite{Flynn}:
\begin{equation}
F_{B_d}=175\pm 25~MeV \qquad
B_{B_d}=1.31\pm 0.03
\end{equation}
This result for $F_{B_d}$ is compatible with the results obtained using
QCD sum rules \cite{QCDSF} and the 
 QCD dispersion relations \cite{BGL}. In our numerical analysis 
we will use
$F_{B_d}\sqrt{B_{B_d}}=200\pm 40~MeV$. The experimental situation on
$(\Delta M)_d$ has been summarized by Gibbons \cite{Gibbons}
 and is given in table 2. For $\tau(B_d)=1.55~ps$ one has then 
$x_d= 0.72\pm 0.03$.

\noindent
{\bf Step 5:}

\noindent
{}The measurement of $B^0_s-\bar B^0_s$ mixing parametrized by $(\Delta M)_s$
together with $(\Delta M)_d$  allows to determine $R_t$ in a different
way. Setting $(\Delta M)^{max}_d= 0.482/ps$ and 
$|V_{ts}/V_{cb}|^{max}=0.993$ (see tables 2 and 3) I find 
a useful formula:
\begin{equation}\label{107b}
(R_t)_{max} = 1.0 \cdot \xi \sqrt{\frac{10.2/ps}{(\Delta M)_s}}
\quad
\xi = 
\frac{F_{B_s} \sqrt{B_{B_s}}}{F_{B_d} \sqrt{B_{B_d}}} 
\end{equation}
where $\xi=1$ in the  SU(3)--flavour limit.
Note that $\mt$ and $|V_{cb}|$ dependences have been eliminated this way
 and that $\xi$ should  
contain smaller theoretical
uncertainties than the hadronic matrix elements in $(\Delta M)_d$ and 
$(\Delta M)_s$ separately.  

The most recent values relevant for (\ref{107b}) are:
\begin{equation}\label{107c}
(\Delta M)_s > 9.2/ ps
\qquad
\xi=1.15\pm 0.05
\end{equation}
The first number is the improved lower bound quoted in \cite{Gibbons}
based in particular on ALEPH and DELPHI results.
The second number comes from quenched lattice calculations summarized
by Flynn \cite{Flynn}.
A similar result has been obtained using QCD sum rules \cite{NAR}.
On the other hand another recent quenched lattice calculation \cite{Soni}
not included in (\ref{107c}) finds
$\xi\approx 1.3 $. Moreover one expects that unquenching will increase
the value of $\xi$ in (\ref{107c}) by roughly 10\% so that values as
high as $\xi=1.25-1.30$ are certainly possible even from Flynn's point of
view. For such high values of $\xi$
the lower bound on $(\Delta M)_s$ in (\ref{107c}) implies $R_t\le 1.37$
which as we will see is similar to the bound obtained on the basis of
the first four steps alone. On the other hand for $\xi=1.15$ one finds
$R_t \le 1.21 $ which puts an additional constraint on the unitarity
triangle cutting lower values of $\bar\varrho$ and higher values 
of $|V_{td}|$. In
view of remaining large uncertainties in $\xi$ we will not use the
constraint from $(\Delta M)_s$ below.

\subsection{Numerical Results}
\subsubsection{Input Parameters}
 The input parameters needed to perform the
standard analysis are given in table 2. The details on the chosen
ranges of $\left| V_{cb} \right|$ and $\left| V_{ub}/V_{cb} \right|$
can be found in \cite{Gibbons}.
Clearly during the last two years there has been a considerable progress
done by experimentalists and theorists in the extraction of
$\left| V_{cb} \right|$ from exclusive and inclusive decays. In
particular I would like to mention important papers by
Shifman, Uraltsev and Vainshtein \cite{SUV},
Neubert \cite{Neubert} and 
Ball, Benecke and Braun \cite{Braun}
on the basis of which one is entitled to use the value given in table 2.
The value for $\left| V_{cb} \right|$ in table 2 incorporates also
inclusive measurements and is slightly higher than the one in
\cite{Gibbons}.
In the case of $\left| {V_{ub}}/{V_{cb}} \right|$
the situation is much worse but progress in the next few years is to be
expected in particular due to new information coming from
exclusive decays \cite{CLEOU,Gibbons}, the inclusive semileptonic
$b\to u$ rate \cite{SUV,Braun,URAL} and the hadronic energy spectrum
in $ B\to X_u e \bar\nu$ \cite{GR96}.

Next it is important to stress that the discovery of the top quark 
by CDF and D0 and its impressive mass measurement summarized by
Tipton \cite{Tipton} had an important impact on
the field of rare decays and CP violation reducing considerably one
potential uncertainty. However
the parameter $\mt$, the top quark mass, used in weak decays is not
equal to the one measured by CDF and D0.
The latter experiments extract the so-called pole mass,
whereas in all  NLO calculations in weak decays  $\mt$ refers
to the running current top quark mass normalized at $\mu=\mt$:
$\mtb(\mt)$. 
For $\mt={\cal O}(170~GeV)$, $\mtb(\mt)$ is typically
by $8~GeV$ smaller than $\mt^{Pole}$. This difference matters already
because the most recent pole mass value has a very small error,
 $175\pm 6~GeV$ \cite{Tipton}, implying
$167\pm 6~GeV$ for $\mtb(\mt)$.
 In this review we will often denote this mass by $\mt$.
Finally the value of $\Lms^{(4)}$ in table 2 corresponds to
$\alpha_s(\mz)=0.118\pm0.005$ with the error slightly larger than
given by Schmelling at this conference.

\begin{table}[thb]
\begin{center}
\begin{tabular}{|c||c|c|}
\hline
{\bf Quantity} & {\bf Central} & {\bf Error} \\
\hline
$|V_{cb}|$ & 0.040 & $\pm 0.003$ \\
$|V_{ub}/V_{cb}|$ & 0.080 & $\pm 0.020$ \\
$B_K$ & 0.75 & $\pm 0.15$ \\
$\sqrt{B_d} F_{B_{d}}$ & $200\mev$ & $\pm 40\mev$ \\
$\sqrt{B_s} F_{B_{s}}$ & $240\mev$ & $\pm 40\mev$ \\
$\mt$ & $167\gev$ & $\pm 6\gev$ \\
$(\Delta M)_d$ & $0.464~ps^{-1}$ & $\pm 0.018~ps^{-1}$ \\
$\Lms^{(4)}$ & $325\mev$ & $\pm 80\mev$ \\
\hline
\end{tabular}
\caption[]{Collection of input parameters.\label{tab:inputparams}}
\end{center}
\end{table}

\subsubsection{Output of the Standard Analysis}
The output of the standard analysis depends to some extent on the
error analysis. This should be always remembered in view of the fact
that different authors use different procedures. In order to illustrate
this  I show in table 3 the results for various quantities of interest
using two types of the error analyses:

\begin{itemize}
\item
Scanning: Both the experimentally measured numbers and the theoretical input
parameters are scanned independently within the errors given in
table~\ref{tab:inputparams}. 
\item
Gaussian: The experimentally measured numbers and the theoretical input 
parameters are used with Gaussian errors.
\end{itemize}
Clearly the "scanning" method is conservative. On the other
hand using Gaussian distributions for theoretical input parameters
can certainly be questioned. 
I think that
at present the conservative "scanning" method should be preferred.
In the future however when data and theory improve, it would be useful to  
find a less conservative estimate which most probably will give errors
somewhere inbetween these two error estimates. 
The analysis discussed here is based on 
\cite{BJL96b}.

\begin{table}
\begin{center}
\begin{tabular}{|c||c||c|}\hline
{\bf Quantity} & {\bf Scanning} & {\bf Gaussian} \\ \hline
$\mid V_{td}\mid/10^{-3}$ &$6.9 - 11.3$ &$ 8.6\pm 1.1$ \\ \hline
$\mid V_{ts}/V_{cb}\mid$ &$0.959 - 0.993$ &$0.976\pm 0.010$  \\ \hline
$\mid V_{td}/V_{ts}\mid$ &$0.16 - 0.31$ &$0.213\pm 0.034$  \\ \hline
$\sin(2\beta)$ &$0.36 - 0.80$ &$ 0.66\pm0.13 $ \\ \hline
$\sin(2\alpha)$ &$-0.76 - 1.0$ &$ 0.11\pm 0.55 $ \\ \hline
$\sin(\gamma)$ &$0.66 - 1.0 $ &$ 0.88\pm0.10 $ \\ \hline
$\IM \lambda_t/10^{-4}$ &$0.86 - 1.71 $ &$ 1.29\pm 0.22 $ \\ \hline
$(\Delta M)_s ps$ &$ 8.0 - 25.4$ &$ 15.2 \pm 5.5 $ \\ \hline
\end{tabular}
\caption[]{Output of the Standard Analysis. 
 $\lambda_t=V^*_{ts} V_{td}$.\label{TAB2}}
\end{center}
\end{table}

\begin{figure}
\centerline{
\psfig{figure=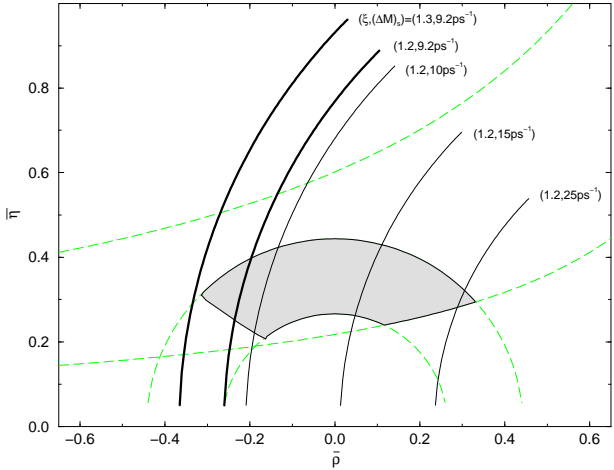,height=3.25in,angle=270}}
\caption[]{Unitarity Triangle 1996.
\label{fig:utdata}}
\end{figure}

In fig. 4  we show the range for the upper
corner A of the UT. The solid thin lines correspond to $R_t^{max}$ from 
(\ref{107b})
using $\xi=1.20$ and $(\Delta M)_s=10/ps,~15/ps$ and $25/ps$, respectively.
The allowed region has a typical "banana" shape which can be found
in many other analyses \cite{BLO,ciuchini:95,HNb,ALUT,PP,PW}. The size of
the banana and its position depends on the used input
parameters and on the error analysis which varies from paper
to paper. The results in fig. 4 correspond to the
scanning method.
We show also the impact of the experimental bound $(\Delta M)_s>9.2/ps$
with $\xi=1.20$ and the corresponding bound for $\xi=1.30$. In view
of the remaining uncertainty in $\xi$, in particular due to quenching,
 this bound has not 
been used in
obtaining the results in table 3. It is evident however that $B^0_s-\bar
B^0_s$ mixing will have a considerable impact on the unitarity triangle
when the value of $\xi$ will be better know and the data improves.
This is very desirable because as seen in fig. 4 our knowledge of
the unitarity triangle is still rather poor. Similarly the uncertainty
in the predicted value of $(\Delta M)_s$ using $\sqrt{B_s} F_{B_{s}}$
of table 1 
is large with central values
around $15/ps$. 
\section{ $\varepsilon'/\varepsilon$}

The measurement of $\varepsilon'/\varepsilon$ at the $10^{-4}$ level
remains as one of the important targets of contemporary particle
physics. A non-vanishing value of this ratio would give the first
signal for the direct CP violation ruling out the superweak models.
The experimental situation on Re($\varepsilon'/\varepsilon$) is unclear
at present. While the result of NA31 collaboration at CERN \cite{barr:93}
with
Re$(\varepsilon'/\varepsilon) = (23 \pm 7)\cdot 10^{-4}$ 
clearly indicates direct CP violation, the value of E731 at Fermilab
\cite{gibbons:93},
Re$(\varepsilon'/\varepsilon) = (7.4 \pm 5.9)\cdot 10^{-4}$,
is compatible with superweak theories
\cite{WO1} in which $\varepsilon'/\varepsilon = 0$.
 Hopefully, in about two years the experimental situation concerning
$\varepsilon'/\varepsilon$ will be clarified through the improved
measurements by the two collaborations at the $10^{-4}$ level and by
the KLOE experiment at  DA$\Phi$NE.

In the standard
model $\varepsilon'/\varepsilon $ is governed by QCD penguins and
electroweak (EW) penguins. In spite of being suppressed by
$\alpha/\alpha_s$ relative to QCD penguin contributions, the
electroweak penguin contributions have to be included because of the
additional enhancement factor ${\rm Re}A_0/{\rm Re}A_2=22$ relative
to QCD penguins. With increasing $\mt$ the EW penguins become
increasingly important \cite{flynn:89,buchallaetal:90}, and entering
$\varepsilon'/\varepsilon$ with the opposite sign to QCD penguins
suppress this ratio for large $\mt$. For $\mt\approx 200\,\gev$ the ratio
can even be zero \cite{buchallaetal:90}.  Because of this strong
cancellation between two dominant contributions and due to uncertainties
related to hadronic matrix elements of the relevant local operators, a
precise prediction of $\varepsilon'/\varepsilon$ is not possible at
present.

In spite of all these difficulties, a considerable progress has been
made in this decade to calculate $\varepsilon'/\varepsilon$. First of
all the complete next-to-leading order (NLO) effective hamiltonians for
$\Delta S=1$ \cite{BJLW,ROMA2}, $\Delta S=2$
\cite{BJW,HNa,HNb} and $\Delta B=2$ \cite{BJW} are now available so that a
complete NLO analysis of $\varepsilon'/\varepsilon$ including
constraints from the observed indirect CP violation ($\varepsilon_K$)
and the $B^0_d-\bar B^0_d$ mixing ($(\Delta M)_d$) is possible. The improved
determination of the $V_{ub}$ and $V_{cb}$ elements of the CKM matrix
\cite{Gibbons}, and in particular the determination of the top quark mass
$\mt$ \cite{Tipton} had of course also an important impact on
$\varepsilon'/\varepsilon$. The main remaining theoretical
uncertainties in this ratio are then the poorly known hadronic matrix
elements of the relevant QCD penguin and electroweak penguin operators
represented by  two important B-factors
($B_6=$ the dominant QCD penguin $Q_6$ and $B_8=$ the dominant electroweak
 penguin $Q_8$), the values of the ${\rm V_{CKM}}$ factors
 and as stressed in \cite{BJLW} the value of $\ms$ and
$\Lambda_{\overline{MS}}$.

An analytic formula for
$\varepsilon'/\varepsilon$ which exhibits all these uncertainties
can be found in \cite{BLAU,BJL96a}. 
A very simplified version of this formula is given as follows
\begin{eqnarray}
\frac{\varepsilon'}{\varepsilon}&=&11\cdot 10^{-4}\left[ 
\frac{\eta\lambda^5 A^2}{1.3\cdot 10^{-4}}\right]
\left[\frac{140~MeV}{ m_s(2~GeV)} \right]^2 
 \nonumber \\
& & \times
\left[\frac{\Lms^{(4)}}{300~MeV} \right]^{0.8}
 [B_6-Z(x_t)B_8]
\label{7e}
\end{eqnarray} 
where $Z(x_t)\approx 0.18 (\mt/\mw)^{1.86}$ and equals unity for 
$\mt\approx 200~GeV$. This simplified formula should not be used
for any serious numerical analysis.

Concerning the values of $B_6$ and $B_8$ one has $B_6=B_8=1$ in the 
vacuum insertion estimate of the hadronic matrix elements in question.
The same result is found in the large $N$ limit
\cite{bardeen:87,burasgerard:87}. Also lattice calculations give
similar results: $B_6=1.0 \pm 0.2$ \cite{kilcup:91,sharpe:91} and $B_8
= 1.0 \pm 0.2$ \cite{kilcup:91,sharpe:91,bernardsoni:89,francoetal:89},
$B_8= 0.81(1)$ \cite{gupta2:96}. These are the values used in
\cite{BJLW,ciuchini:95,BBL,BJL96a}.
In the chiral quark model one finds \cite{bertolinietal:95}: $B_6=1.0\pm0.4$,
$B_8=2.2\pm1.5$ and generally $B_8>B_6$. On the other hand the Dortmund group
\cite{heinrichetal:92} advocates  $B_6>B_8$.
>From \cite{heinrichetal:92} $B_6=1.3$ and $B_8=0.7$ can be extracted.

What about $\ms$? 
 The most recent results of QCD sum
rule (QCDSR) calculations \cite{jaminmuenz:95,chetyrkinetal:95,narison:95}
obtained at $\mu=1\gev$ correspond to $\ms(2\gev)=145\pm20\mev$.
The lattice calculation of \cite{alltonetal:94} finds
$\ms(2~GeV)=128\pm18~MeV$, 
in rather good agreement with the QCDSR result. This summer a new lattice
result has been presented by Gupta and Bhattacharya \cite{gupta:96}. They
find $\ms(2\gev)=90\pm20\mev$ which is on the low side of all strange mass 
determinations. Moreover they find that unquenching lowers further
the values of $\ms(2\gev)$ to $70\pm15\mev$. 
Similar results are found by the FNAL 
group \cite{FNALL}.
The situation with the strange quark mass is therefore unclear at present
and hopefully will be clarified soon.

The most recent analysis of \cite{BJL96a} using input parameters of table 1,
$B_6=1.0\pm 0.2$, $B_8=1.0\pm 0.2$ and 
$m_s(2~GeV)=129\pm17\mev$ finds
\begin{equation}
-1.2 \cdot 10^{-4} \le \epe \le 16.0 \cdot 10^{-4}
\label{eq:eperangenew}
\end{equation}
and
\begin{equation}
\epe= ( 3.6\pm 3.4) \cdot 10^{-4}
\label{eq:eperangefinal}
\end{equation}
for the "scanning" method and the "gaussian" method respectively.

The result in (\ref{eq:eperangefinal}) agrees rather well with
the 1995 analysis of the Rome group \cite{ciuchini:95} which finds
 $\varepsilon'/\varepsilon=(3.1\pm 2.5)\cdot 10^{-4}$.
On the other hand the range in (\ref{eq:eperangenew}) shows that for
particular choices of the input parameters, values for $\epe$ as high as
$16\cdot 10^{-4}$ cannot be excluded at present. Such high values are
found if simultaneously  $\vub=0.10$, $B_6=1.2$, $B_8=0.8$, $B_K=0.6$,
$\ms(2\gev)=110 MeV$ and $\Lms^{(4)}=405\mev$ 
are chosen. It is however evident from  the comparision of
(\ref{eq:eperangenew}) and (\ref{eq:eperangefinal})  that such 
high values of $\epe$ and generally values above $10^{-3}$ 
are very improbable.

The authors of \cite{bertolinietal:95} calculating the $B_i$ factors
in the chiral quark model find using the scanning method
 a rather large range $-50 \cdot 10^{-4}\le \epe \le 14 \cdot 10^{-4}$.
 In particular they find in contrast to
\cite{BJLW,ciuchini:95,BBL,BJL96a} 
that negative values for $\epe$ as large as $-5\cdot
10^{-3}$ are possible. 
The Dortmund group
\cite{heinrichetal:92} advocating on the other hand $B_6>B_8$ finds
$\epe=(9.9\pm 4.1)\cdot 10^{-4}$ for $m_s(2~GeV)=130~MeV$.

The situation with $\epe$ in the standard model may however change
 if the value for $m_s$ is as low as found in \cite{gupta:96,FNALL}.
Using 
$\ms(2\gev)=85\pm17\mev$ one finds \cite{BJL96a}
\begin{equation}
0 \le \epe \le 43.0 \cdot 10^{-4}
\label{eq:eperangenewa}
\end{equation}
and
\begin{equation}
\epe= ( 10.4\pm 8.3) \cdot 10^{-4}
\label{eq:eperangefinala}
\end{equation}
for the "scanning" method and the "gaussian" method respectively.
We observe that the "gaussian" result agrees well with the E731
value and
as stressed in \cite{BJL96a} the decrease of $\ms$
 with $\ms(2\gev)\geq 85~MeV$ alone is insufficient to bring 
the standard model to agree with
the NA31 result. However for $B_6>B_8$, sufficiently large values of
$|V_{ub}/V_{cb}|$ and $\Lms$ and small values of $\ms$, the values
of $\epe$ in the standard model can be as large as $(2-4)\cdot 10^{-3}$
and consistent with the NA31 result.
\section{Radiative and Rare B Decays}
\subsection {$B\to X_s\gamma$} 
The rare decay $B\to X_s\gamma$ plays an important role in the
present day phenomenology. It originates from the magnetic 
$\gamma$-penguins of (\ref{O6}).
The perturbative QCD effects 
 are very important in this decay.
They are known
\cite{Bert,Desh} to enhance $B\to X_s\gamma$ in 
the SM by 2--3
times, depending on the top quark mass. Since the first analyses
in \cite{Bert,Desh} a lot of progress has been made in calculating
the QCD effects beginning with the work in \cite{Grin,Odon}. We will
briefly summarize the present status.

A peculiar feature of the renormalization group analysis 
in $B\to X_s\gamma$ is that the mixing under infinite renormalization 
between
the set $(Q_1...Q_6)$ and the operators $(Q_{7\gamma},Q_{8G})$ vanishes
at the one-loop level. Consequently in order to calculate 
the coefficients
$C_{7\gamma}(\mu)$ and $C_{8G}(\mu)$ in the LO
approximation, two-loop calculations of ${\cal{O}}(e g^2_s)$ 
and ${\cal{O}}(g^3_s)$
are necessary. The corresponding NLO analysis requires the evaluation
of the mixing in question at the three-loop level. 

Until recently $B\to X_s\gamma$ has been known
only in the leading logarithmic approximation 
 \cite{LLO,Mis:94}.
However this summer the NLO corrections have been finally completed.
It was a joint effort of many groups.
 The two-loop
mixing involving the operators
$Q_1.....Q_6$ and the two-loop mixing
in the sector $(Q_{7\gamma},Q_{8G})$ has been calculated in 
\cite{ALTA,BW,BJLW1,BJLW,ROMA1,ROMA2} 
and \cite{MisMu:94} respectively. 
 The $O(\alpha_s)$
corrections to $C_{7\gamma}(M_W)$ and $C_{8G}(M_W)$ have been calculated
in \cite{Yao1}. One-loop matrix elements 
$\langle s\gamma {\rm gluon}|Q_i| b\rangle$ have been calculated in 
\cite{AG2,Pott}. The very difficult two-loop corrections to 
$\langle s\gamma |Q_i| b\rangle$ have been presented in \cite{GREUB}.
Finally after a heroic effort  the three loop mixing between
the set $(Q_1...Q_6)$ and the operators $(Q_{7\gamma},Q_{8G})$
 has been completed this summer and presented by Misiak at this
conference \cite{CZMM}.

The leading logarithmic calculations 
   can be summarized in a compact form
as follows \cite{AG1,BMMP:94}.
\begin{equation}\label{main}
\frac{Br(B \to X_s \gamma)}{Br(B \to X_c e \bar{\nu}_e)}
 =  \frac{|V_{ts}^* V_{tb}|^2}{|V_{cb}|^2} 
\frac{6 \alpha_{QED}}{\pi f(z)} |C^{(0)eff}_{7\gamma}(\mu)|^2
\end{equation}
 where
$C^{(0)eff}_{7\gamma}(\mu)$ is the effective coefficient 
for which an analytic expression can be found in \cite{BMMP:94},
 $z = {m_c}/{m_b}$, and
$f(z)$ is the phase space factor in the semileptonic
b-decay. 
The expression given above is based on the
spectator model corrected for short-distance QCD effects. 
The ${\cal O}(1/m_b^2)$ corrections to this result
have been studied in
\cite{Chay,Bj,Bigi} and found to be at the level of  
 a few percent.

A critical analysis of theoretical and
experimental
uncertainties present in the prediction for Br(\Bsg) based on the
formula (\ref{main}) has been made in \cite{BMMP:94} giving
\begin{equation}
Br(B \to X_s\gamma)_{TH} = (2.8 \pm 0.8) \times 10^{-4}
\label{theo}
\end{equation}
where the error is dominated by the uncertainty in 
the choice of the renormalization scale
$m_b/2<\mu<2 m_b$ as first stressed by Ali and Greub \cite{AG1} and confirmed
in \cite{BMMP:94}.
This large $\mu$ uncertainty of $\pm 20\%$ has been recently reduced down to
$\pm 6\%$ \cite{GREUB,CZMM} through the complete NLO calculations 
mentioned above.
Including additional uncertainties in the values of $\alpha_s$, $z=m_c/m_b$,
$\mt$ and Br($B \to X_c e \bar{\nu}_e$) the preliminary result reads
\cite{GREUB,CZMM}	
\begin{equation}
Br(B \to X_s\gamma)_{TH} = (3.3 \pm 0.5) \times 10^{-4}
\label{t96}
\end{equation}

The result in (\ref{t96}) should be compared with the CLEO 
1994 measurement
\cite{CLEO2}:
\begin{equation}
Br(B \to X_s\gamma) = (2.32 \pm 0.57 \pm 0.35) \times 10^{-4}
\label{incl}
\end{equation}
where the first error is statistical and the second is systematic.
At this conference a 90\% C.L. upper bound $5.4\cdot 10^{-4}$, consistent
with (\ref{incl}) has been presented by DELPHI (PA01-051).

The result in (\ref{incl}) is compatible with (\ref{t96}) although 
the theoretical and experimental errors should be decreased in
the future in order to reach a definite conclusion and to see
whether some contributions beyond the standard model are required.
In any case the agreement of the
theory with data is consistent with the large QCD enhancement
of \Bsg. 
The result in (\ref{t96}) should also be compared with a partial
NLO analysis of
\cite{Ciu:94} which implied 
$Br(B\to X_s\gamma)=(1.9\pm 0.2\pm 0.5)\cdot 10^{-4}$.
Indeed partial NLO analyses can be sometimes
misleading.

In order to find possible implications for new physics such as represented
by Two-Higgs-Doublet Model (2HDM) or the Minimal Supersymmetric Standard
Model (MSSM) the ${\cal O} (\alpha)$ corrections to $C_{7\gamma}(\mw)$
in these models have to be calculated. In spite of the absence of such
calculations some rough statements can be made already now however.
>From leading order analyses one has an intersting lower bound on the mass of
the charged Higgs in the
most popular two Higgs doublet model. At $95\%~C.L.$
CLEO \cite{CLEO2} finds $m_{H^{\pm}}> 250~GeV$.
The enhancement of the theoretical branching ratio through NLO corrections
should increase this bound to roughly $500 \gev$. In MSSM where also
important chargino and stop contributions are present, both an enhancement
and an suppression relative to the standard model result are possible.
A suppression could even be necessary if the central value in (\ref{incl})
will not be substantially modified by future improved measurements.
This would put some restrictions on the values of the free parameters in
MSSM. Clearly the chapter on \Bsg is far from being closed and we should
look forward to the coming years which should be very exciting in this area.
 
In 1993
CLEO reported \cite{CLEO} 
$Br(B \to K^* \gamma) = (4.5 \pm 1.5 \pm 0.9) \times 10^{-5}.$
The corresponding 1996 value has a substantially smaller error
(PA05-093):
\begin{equation}
Br(B \to K^* \gamma) = (4.2 \pm 0.8 \pm 0.6) \times 10^{-5}
\label{excl}
\end{equation}
implying an improved measurement of $R_{K^*}$:

\begin{equation}
R_{K^*}=\frac{\Gamma (B \to K^* \gamma)}{\Gamma(B \to X_s \gamma)}
 = 0.181 \pm 0.068
\label{K*}
\end{equation}

This result puts some constraints on various formfactor models
listed in (PA05-093). There one can also find 90\% C.L. bounds 
$Br (B^0\to\rho^0\gamma)< 3.9\cdot 10^{-5}$,
$Br (B^0\to\omega\gamma)< 1.3\cdot 10^{-5}$,
$Br (B^-\to\rho^-\gamma)< 1.1\cdot 10^{-5}$.
Combined with (\ref{excl}) these
bounds imply $\vtd/\vts\le 0.45-0.56$ where the uncertainty in
the bound reflects the model dependence. Clearly this bound is
still higher by a factor of two than the value obtained in the
standard analysis of section 3. Next bounds from DELPHI (PA01-051)
$Br (B^0_d\to K^*\gamma)< 2.1\cdot 10^{-4}$,
$Br (B^0_s\to\phi\gamma)< 7.0\cdot 10^{-4}$ should
be mentioned here.

As advertised at this conference by 
Browder, Kagan and Kagan, CLEO II
should be able to discover $b \to s {\rm gluon}$ transition soon.
In view of this, theorists should sharpen their tools. More
on \Bsg, in particular on the photon spectrum and the determination
of $\vtd/\vts$ from $B\to X_{s,d}\gamma$, can be found in
\cite{ALUT,ALIB}.

\subsection{$B\to X_s \mu^+\mu^-$ }
The rare decay $B\to X_s \mu^+\mu^-$ is dominated by the electroweak
penguin of fig. 1f and receives small contributions from 
box diagrams and magnetic penguins. 
The presence of electroweak penguin diagrams 
implies a strong $\mt$ dependence of $Br(B\to X_s \mu^+\mu^-)$ as
stressed in particular by  Hou et al. \cite{HWS:87}.
There is a vast literature on this decay which can be found e.g. in
\cite{ALUT,ALUT96} and \cite{BuMu:94}. 
Here I will consider only certain
aspects.

The presence of $c\bar c$ resonances $\psi$, $\psi'$ in the
$\mu^+\mu^-$ channel complicates the theoretical analysis as
discussed in detail in \cite{ALUT,ALUT96}. The non-resonant part can 
however be calculated with high confidence.
The QCD corrections to this part have been calculated 
 over the last years with increasing precision by several
groups \cite{GSW:89,GDSN:89,CRV:91,Mis:94} culminating in two complete
next-to-leading QCD calculations 
\cite{Mis:94,BuMu:94} which agree with each other.
An important gain due to these NLO calculations is a considerable
reduction in the $\mu$-dependence of the resulting branching ratio.
Whereas in LO an uncertainty as large as $\pm 20\%$ can be found,
it is reduced as shown in \cite{BuMu:94} below $\pm 5\%$ after the
inclusion of NLO corrections. 
For central values of parameters the NLO branching ratio turns out to
be enhanced by $10\%$ over its LO value. Choosing 
$Br(B \to X_c e \bar{\nu}_e)=10.4\%$ the final result for the
non-resonant part can be well approximated by:
\begin{eqnarray}
\lefteqn{Br(B\to X_s \mu^+\mu^-)_{NR}=}
\nonumber \\
&&
6.2\cdot 10^{-6} 
\left [\frac{|V_{ts}|}{|V_{cb}|} \right ]^2 
\left [\frac{(\mtb(\mt)}{170\gev} \right ]^2 
\label{MUMU}
\end{eqnarray}

A more detailed analysis gives \cite{BAER}
\begin{equation}\label{MUR}
Br(B\to X_s \mu^+\mu^-)_{NR}=
(5.7\pm 0.9)\cdot 10^{-6}
\end{equation}
with a similar result in \cite{ALUT96}.
This should be compared with the most recent upper bound from 
D0 \cite{DARIA}:
$(3.2)\cdot 10^{-5}$ which improves the 1991 bound of UA1 by roughly a 
factor of two. It is exciting that the experimental bound is
only by a factor of five above the standard model expectations.
D0 should be able to measure this branching ratio during the Run II
at Tevatron.

Clearly the calculation of $Br(B\to X_s \mu^+\mu^-)$ is only a small
part of the activities present in the literature. The invariant
dilepton mass spectrum, the forward-backward charge asymmetry and 
various lepton polarization asymmetries (in particular in
$B\to X_s\tau^+\tau^-$) should enable a detailed study 
of the dynamics of the standard model and the search for new 
physics beyond it. This has been stressed by Hewett at 
this conference and in \cite{ASYMM,ALUT,ALIB,GLN96} 
where further references can be found. 

It should be stressed that the $1/m^2_b$ corrections in this
decay calculated very recently amount to only a few percent \cite{ALUT96}
(see also \cite{PBAER})
as opposed to $10\%$ found earlier \cite{FALK}.

The Standard Model predictions for exclusive channels 
$B_d\to K^* e^+e^-$ and $B_d\to K^* \mu^+\mu^-$ amount to
$(2.3\pm0.9)\cdot 10^{-6}$ and $(1.5\pm 0.6)\cdot 10^{-6}$
respectively \cite{ALUT}. This should be compared with the upper bounds
$1.6\cdot 10^{-5}$ and $2.5\cdot 10^{-5}$ by CLEO and CDF respectively.
The upper bound on the sum of these two channels from DELPHI (PA01-091)
amounts to $2.4 \cdot 10^{-3}$. 
Sensitivity of $3\cdot 10^{-7}$ should be reached for
$B_d\to K^* \mu^+\mu^-$ by CDF during the Run II. The exclusive channels,
although not as clean as the inclusive ones, should also offer  
some insight in the dynamics involved. 

\subsection{ $B\to\mu\bar\mu$ and $B\to X_s\nu\bar\nu$}

$B\to\mu\bar\mu$ and $B\to X_s\nu\bar\nu$ are the theoretically
cleanest decays in the field of rare B-decays.
$B\to\mu\bar\mu$ and $B\to X_s\nu\bar\nu$
 are dominated by the $Z^0$-penguin and box diagrams
involving top quark exchanges.
The NLO corrections to both decays have been calculated in
\cite{BB2}. These calculations reduced
considerably the theoretical uncertainties in the branching ratios
related to the scale $\mu_t$ present in $\mtb(\mu_t)$.
In the case of $B\to\mu\bar\mu$ this reduction is roughly from
$\pm 13\%$ to $\pm 1\%$. In the case of $B\to X_s\nu\bar\nu$
from $\pm$ 10\% to $\pm 1\%$.
Choosing
$\mu_t=m_t$ the final expressions for the branching ratios in question
take a particularly
simple form \cite{BB2}. Updating the input parameters one has:
\begin{eqnarray}
\lefteqn{Br(B_s\to \mu\bar\mu)=
3.4\cdot 10^{-9}\left[\frac{F_{B_s}}{210~MeV}\right]^2}
\nonumber \\
&& \times
\left[\frac{\mtb(\mt)}{170~GeV} \right]^{3.12} 
\left[\frac{\mid V_{ts}\mid}{0.040} \right]^2 
\left[\frac{\tau_{B_s}}{1.6 ps} \right] 
\end{eqnarray}
and
\begin{equation}
Br(B\to X_s\nu\bar\nu)=
4.1\cdot 10^{-5}
\left[\frac{|V_{ts}|}{|V_{cb}|} \right]^2 
\left[\frac{\mtb(\mt)}{170~GeV} \right]^{2.3} 
\end{equation}
where in the last expression we have used 
$Br(B \to X_c e \bar{\nu}_e)=10.4\%$

Scanning the input parameters of table 2 one finds \cite{BJL96b}
\begin{equation}\label{122}
Br(B_s\to \mu\bar\mu)=(3.6 \pm 1.8)\cdot 10^{-9} 
\end{equation}
\begin{equation}\label{123}
Br(B\to X_s \nu\bar\nu)=(3.8 \pm 0.8)\cdot 10^{-5} 
\end{equation}
where in obtaining (\ref{122})
$F_{B_s}=210\pm30\mev$ has been used.

The strong dominance of the internal top exchanges and a short
distance character of the contributions allows 
a clean determination of $\vtd/\vts$
by measuring the ratios $Br(B_d\to \mu\bar\mu)/Br(B_s\to \mu\bar\mu)$
and $Br(B\to X_d \nu\bar\nu)/Br(B\to X_s\nu\bar\nu)$. In particular
the latter determination is very clean as the uncertainties related to
$Br(B \to X_c e \bar{\nu}_e)$ and $\vcb$ cancel in the ratio.
The corresponding determination using $B_{d,s}\to \mu\bar\mu$
suffers from the uncertainty in the ratio $F_{B_d}/F_{B_s}$ which
however should be removed to a large extend by future lattice
calculations.

One of the high-lights of FCNC-1996 is the upper bound (90\% C.L.):
\begin{equation}\label{124}
Br(B\to X_s \nu\bar\nu) < 7.7\cdot 10^{-4} 
\end{equation}
obtained for the first time by ALEPH (PA10-019).
This is only a factor of 20 above the standard model expectation.
Even if the actual measurement of this decay is extremly difficult,
all efforts should be made to measure it. Meanwhile the bound in 
(\ref{124}) puts some constraints on some exotic physics beyond the
standard model \cite{Ligetti}.
The $90 \%$ C.L. bounds $Br(B_d\to K^*\nu\bar\nu)<1\cdot 10^{-3}$
and $Br(B_s\to \phi\nu\bar\nu)<5.4\cdot 10^{-3}$ from DELPHI
(PA01-051) should be compared with ${\cal O}(10^{-5})$ in the
standard model.

The bounds on $B_{s,d}\to l\bar l$ are still
many orders of magnitude away from standard model expectations.
One has $Br(B_s\to\mu\bar\mu)< 8.4\cdot 10^{-6}$ (CDF),
$8.0\cdot 10^{-6}$ (D0,PA07-024), $5.6\cdot 10^{-5}$ (L3) and
$Br(B_d\to\mu\bar\mu)< 1.6\cdot 10^{-6}$ (CDF),
$1.2\cdot 10^{-5}$ (L3) (PA05-046).
The D0 result is really an upper bound 
on $(B_s+B_d)\to \mu\bar\mu$. CDF should reach in Run II the
sensitivity of $1\cdot 10^{-8}$ and $4\cdot 10^{-8}$ for
$B_d\to \mu\bar\mu$ and $B_s\to \mu\bar\mu$ 
respectively \cite{Lewis}.
 It is hoped that these decays will be observed at
LHC-B. The experimental status of $B\to\tau^+\tau^-$ and its
usefulness in tests of the physics beyond the standard model
is discussed in \cite{GLN96}.

\section{Rare K Decays}
\subsection{The Decay $K_L\to\pi^0 \lowercase{e}^+\lowercase{e}^-$}

Whereas in $K\to\pi\pi$ decays the CP violating contribution is
only a tiny part of the full amplitude and the direct CP violation
is expected to be at least by three orders of
magnitude smaller than the indirect CP violation, the corresponding
hierarchies are very different for $K_L\to\pi^0e^+e^-$. At lowest
order in electroweak interactions (one-loop photon penguin,
$Z^0$-penguin and W-box diagrams), this decay takes place only if
CP symmetry is violated. The CP conserving contribution to the
amplitude comes from a two photon exchange, which although of higher
order in $\alpha$ could in principle be sizable. 
The CP
violating part can again be divided into a direct and an indirect one.
The latter is given by the $K_S\to\pi^0e^+e^-$ amplitude times the CP
violating parameter $\varepsilon_K$.

Only the directly CP violating contribution can be calculated reliably
at present.
The other two contributions are unfortunately very uncertain 
and the following ranges can be found in the literature:
\begin{equation}
Br(K_L \to \pi^0 e^+ e^-)_{cons}\approx\left\{ \begin{array}{ll}
(0.3-1.8)\cdot 10^{-12} & \hbox{\cite{cohenetal:93}} \\
     4.0 \cdot 10^{-12} & \hbox{\cite{heiligerseghal:93}} \\
(5 \pm 5)\cdot 10^{-12} & \hbox{\cite{donoghuegabbiani:95}}
\end{array} \right.
\label{eq:BKLtheo}
\end{equation}
and \cite{eckeretal:88,brunoprades:93,heiligerseghal:93,donoghuegabbiani:95}
\begin{equation}
Br(K_L \to \pi^0 e^+ e^-)_{indir}=(1.-5.)\cdot 10^{-12} 
\label{eq:BKLindir2}
\end{equation}

On the other hand there are
practically no theoretical uncertainties in the directly CP violating
part 
because the relevant matrix element 
$\langle\pi^0|(\bar sd)_{V-A}| K_L\rangle$ can be
extracted using isospin symmetry from the well measured decay
$K^+\to\pi^0e^+\nu$. Calculating the relevant box and electroweak
penguin diagrams and including LO \cite{GIL2} and NLO \cite{BLMM} 
corrections one finds
approximately:
\begin{eqnarray}
\lefteqn{Br(K_L \to \pi^0 e^+ e^-)_{dir}=
4.4\cdot 10^{-12}}
\nonumber \\
& & \times
\left [ \frac{\eta}{0.37}\right ]^2
\left [\frac{|V_{cb}|}{0.040} \right ]^4 
\left [\frac{\mtb(\mt)}{170\gev} \right ]^2 
\label{KT}
\end{eqnarray}
Scanning the input parameters of table 2 gives \cite{BJL96b}
\begin{equation}\label{12}
Br(K_L\to\pi^0 e^+ e^-)_{dir}=(4.5 \pm 2.6)\cdot 10^{-12} 
\end{equation}
where the error comes dominantly from the uncertainties in the CKM
parameters. Thus the directly CP violating contribution is comparable
to the other two contributions and could even be dominant.
In order to see whether this is indeed the case  improved estimates of the
other two contributions are necessary.

A much better assessment of the importance of the
indirect CP violation in $K_L\to\pi^0e^+e^-$ will become possible after
a measurement of $Br(K_S\to\pi^0e^+e^-)$.  Bounding the latter
branching ratio below $ 1 \cdot 10^{-9}$ or $ 1 \cdot 10^{-10}$ would
bound the indirect CP violating 
 contribution below $ 3 \cdot 10^{-12}$ and
 $ 3 \cdot 10^{-13}$ respectively. The present bounds: 
$ 1.1 \cdot 10^{-6}$ (NA31) and  $ 3.9 \cdot 10^{-7}$ (E621) are still 
too weak. On the other hand KLOE at DA${\Phi}$NE
could make an important contribution here.  

The present experimental bounds
\begin{equation}
Br(K_L\to\pi^0 e^+ e^-) \leq \left\lbrace \begin{array}{ll}
4.3 \cdot 10^{-9} & \mbox{\cite{harris}} \\
5.5 \cdot 10^{-9} & \mbox{\cite{ohl}} \end{array} \right.
\end{equation}
are still by three orders of magnitude away from the theoretical
expectations in the Standard Model. Yet the prospects of getting the
required sensitivity of order $10^{-11}$--$10^{-12}$ by 1999 are
encouraging \cite{CPRARE}. More details on this decay 
can be found in the recent review by Pich \cite{PICH96}.
\subsection{$K_L\to\pi^0\nu\bar\nu$ and $K^+\to\pi^+\nu\bar\nu$}
$K_L\to\pi^0\nu\bar\nu$ and $K^+\to\pi^+\nu\bar\nu$ are the theoretically
cleanest decays in the field of rare K-decays. 
$K_L\to\pi^0\nu\bar\nu$ is 
dominated by short distance loop diagrams (Z-penguins and box diagrams)
involving the top quark.  $K^+\to\pi^+\nu\bar\nu$ receives
additional sizable contributions from internal charm exchanges.
The great virtue of $K_L\to\pi^0\nu\bar\nu$ is that it proceeds
almost exclusively through direct CP violation \cite{Littenberg} 
and as such is the
cleanest decay to measure this important phenomenon. It also offers
a clean determination of the Wolfenstein parameter $\eta$ and in particular
as we will stress in section 8 offers the cleanest measurement
of $\IM\lambda_t= \IM V^*_{ts} V_{td}$ 
which governs all  CP violating  K-decays. 
$K^+\to\pi^+\nu\bar\nu$ is CP conserving and offers a clean 
determination of $|V_{td}|$. Due to the presence of the charm
contribution and the related $m_c$ dependence it has a small
scale uncertainty absent in $K_L\to\pi^0\nu\bar\nu$.

The next-to-leading QCD corrections to both decays
have been calculated in 
\cite{BB1,BB2,BB3}. These calculations
considerably reduced the theoretical uncertainty
due to the choice of the renormalization scales present in the
leading order expressions \cite{DDG}, in particular in the charm contribution
to $K^+\to\pi^+\nu\bar\nu$. Since the relevant hadronic matrix
elements of the weak currents entering $K\to \pi\nu\bar\nu$
can be related using isospin symmetry to the leading
decay $K^+ \rightarrow \pi^0 e^+ \nu$, the resulting theoretical
expressions for Br( $K_L\to\pi^0\nu\bar\nu$) and Br($K^+\to\pi^+\nu\bar\nu$)
  are only functions of the CKM parameters, the QCD scale
 $\Lms$
 and the
quark masses $\mt$ and $\mc$. The isospin braking corrections calculated in
\cite{MP} reduce the $K^+$ and $K_L$ branching ratios by $10\%$ and $5.6\%$
respectively.
The long distance contributions to
$K^+ \rightarrow \pi^+ \nu \bar{\nu}$ have been
considered in \cite{RS} and found to be very small: a few percent of the
charm contribution to the amplitude at most, which is safely neglegible.
The long distance contributions to $K_L\to\pi^0\nu\bar\nu$ are negligible
as well. 

The explicit expressions for $Br(K^+ \rightarrow \pi^+ \nu \bar{\nu})$ 
and $Br(K_L\to\pi^0\nu\bar\nu)$ can be found in \cite{BBL}. Here we
give approximate expressions in order to exhibit various dependences:

\begin{eqnarray}
\lefteqn{Br(K^+ \rightarrow \pi^+ \nu \bar{\nu})=
0.7\cdot 10^{-10}
\Bigl\{ \left [ \frac{|V_{td}|}{0.010}\right ]^2}
\label{bkpn} \\
&& \times
\left [\frac{\mid V_{cb}\mid}{0.040} \right ]^2
\left [\frac{\mtb(\mt)}{170~\gev} \right ]^{2.3} 
 +\mbox{cc} + \mbox{tc}\Bigr\}
\nonumber
\end{eqnarray}

\begin{eqnarray}
\lefteqn{Br(K_L\to\pi^0\nu\bar\nu)=
2.8\cdot 10^{-11}}
\nonumber \\
&& \times
\left [ \frac{\eta}{0.37}\right ]^2
\left [\frac{\mtb(\mt)}{170~GeV} \right ]^{2.3} 
\left [\frac{\mid V_{cb}\mid}{0.040} \right ]^4 
\label{bklpn}
\end{eqnarray}
where in (\ref{bkpn}) we have shown explicitly only the pure top
contribution.

The main impact of NLO is the reduction of
scale uncertainties 
in the case of $K^+ \rightarrow \pi^+ \nu \bar{\nu}$ roughly from
$\pm 22\%$ to $\pm 7\%$. and in the case of $K_L\to \pi^0\nu\bar\nu$
from $\pm 10\%$ to $\pm 1\%$.
The reduction of the scale uncertainty in 
$Br(K^+ \rightarrow \pi^+ \nu \bar{\nu})$ corresponds to the reduction
in the uncertainty in the determination of $|V_{td}|$ from $\pm 14\%$ to 
$\pm 4\%$.
Scanning the input parameters of table 2 one finds \cite{BJL96b}:
\begin{eqnarray}
Br(K^+ \rightarrow \pi^+ \nu \bar{\nu})=
(9.1\pm 3.2)\cdot 10^{-11}
&&
\nonumber \\
Br(K_L\to\pi^0\nu\bar\nu)=(2.8\pm 1.7)\cdot 10^{-11}
&&
\end{eqnarray}
where the errors come dominantly from the uncertainties in the
CKM parameters. 
The present experimental bound on $Br(K^+ \to \pi^+ \nu \bar\nu)$
is $2.4 \cdot 10^{-9}$ \cite{Adler95}. A new bound $ 2 \cdot 10^{-10}$ for 
this decay  is expected from E787 at AGS in Brookhaven in 1997.
In view of the clean character of this decay a measurement of its
branching ratio at this level would signal the presence of physics
beyond the standard model. Further experimental improvements for
this branching ratio are discussed  
in \cite{Cooper}.
The present upper bound on $Br(K_L\to \pi^0\nu\bar\nu)$ from
FNAL experiment E731 \cite{WEAVER} is $5.8 \cdot 10^{-5}$. 
FNAL-E799 expects to reach
the accuracy ${\cal O}(10^{-8})$ and a very interesting new proposal 
 AGS2000 \cite{AGS2000} 
expects to reach the single event sensitivity $2\cdot 10^{-12}$
allowing a $10\%$ measurement of the expected branching ratio. 
It is hoped that also JNAF(CEBAF), KAMI and KEK will make efforts
to measure this gold-plated  decay. Such measurements will also put
constraints on the physics beyond the standard model \cite{Sher}.

\subsection{$K_L\to \mu\bar\mu$}
The rare decay $K_L\to\mu\bar\mu$ is CP conserving 
and in addition to its
short-distance part, given by Z-penguins and box diagrams, receives 
important contributions from the
two-photon intermediate state, which are difficult to calculate
reliably \cite{gengng:90,belangergeng:91,ko:92,Singer,eeg}.
This latter fact is rather unfortunate because the
short-distance part is, similarly to $K\to\pi\nu\bar\nu$, free of hadronic
uncertainties and if extracted from the existing data would give a useful
determination of the Wolfenstein parameter $\varrho$.
 The separation
of the short-distance piece from the long-distance piece in the measured
rate is very difficult however.

The NLO corrections to the short distance part
have been calculated in \cite{BB1,BB2,BB3}. This calculation
reduced the theoretical uncertainty
due to the choice of the renormalization scales present in the
leading order expressions from $\pm 24\%$ to $\pm 10\%$. The
remaining scale uncertainty which is larger than in 
$K^+ \rightarrow \pi^+ \nu \bar{\nu}$ is related to a particular
feature of the perturbative expansion in this decay \cite{BB3}.
An approximate expression for the short distance part is given as
follows:
\begin{eqnarray}
\lefteqn{Br(K_L\to\mu\bar\mu)_{SD} =
0.9\cdot 10^{-9}}
\nonumber \\
&& \times
\left (1.2 - \bar\varrho \right )^2
\left [\frac{\mtb(\mt)}{170~GeV} \right ]^{3.1} 
\left [\frac{\mid V_{cb}\mid}{0.040} \right ]^4 
\label{bkmumu}
\end{eqnarray}
Scanning the input parameters of table 2 gives:
\begin{equation}\label{BSD}
Br(K_L\to\mu\bar\mu)_{SD}=
(1.3\pm 0.6)\cdot 10^{-9}
\end{equation}
where the error comes dominantly from the uncertainties in the
CKM parameters. 

Now the full branching ratio can be written generally as follows:
\begin{eqnarray}
&& Br(K_L\to\mu\bar\mu)= |\RE A|^2+|\IM A|^2
\nonumber \\
&&
\RE A = A_{SD}+A_{LD}
\label{BSDa}
\end{eqnarray}
with $\RE A$ and $\IM A$ denoting the dispersive and absorptive contributions
respectively. The absorptive contribution can be calculated using the
data for $K_L\to \gamma\gamma$ and is known under the name of the unitarity
bound \cite{Unitary}. 
One finds $(6.81\pm 0.32)\cdot 10^{-9}$ which is very close to the
experimental measurements
\begin{eqnarray}
\lefteqn{Br(K_L\to \bar\mu\mu) =}
\\
&& =
\left\lbrace \begin{array}{ll}
(6.86\pm0.37)\cdot 10^{-9} & \mbox{BNL 791 \cite{PRINZ}} \\
(7.9\pm 0.6 \pm 0.3)\cdot 10^{-9} & 
\mbox{KEK 137 \cite{Akagi}} \end{array} \right.
\nonumber
\end{eqnarray}
which give the world average:
\begin{equation}\label{princ}
Br(K_L\to \bar\mu\mu) = (7.1\pm 0.3)\cdot 10^{-9}
\end{equation}
The accuracy of this result is impressive $(\pm 4\%)$. It will be
reduced to $(\pm 1\%)$ at BNL in the next years.

The BNL791 group using their data and the unitarity bound extracts
$|\RE A|^2\le 0.6\cdot 10^{-9}$ at $90\%$ C.L. This is lower than the
short distance prediction in (\ref{BSD}). Unfortunately in order to use
this result for the determination of $\varrho$ the long distance dispersive 
part $A_{LD}$ resulting from the intermediate off-shell two photon states 
should be known.
The present estimates of $A_{LD}$ are too uncertain to obtain a useful
information on $\varrho$. It is believed that the measurement of
$Br(K_L\to e\bar e \mu\bar\mu)$ should help in estimating this part.
The present result $(2.9+6.7-2.4)\cdot 10^{-9}$ from E799 
should therefore be improved.

More details on this decay can be found in 
\cite{PRINZ,BB3,CPRARE,PICH96,Singer}.
More promising from theoretical point of view is the parity-violating
asymmetry in $K^+\to \pi^+\mu^+\mu^-$ \cite{GENG,BB5}.          
Finally the longitudinal
polarization in this decay is rather sensitive to contributions
beyond the standard model \cite{EKPICH}.

\subsection{$K_L\to \mu e$ and $K\to \pi\mu e$}
 These decays are forbidden in the standard model. Their occurence
would signal the violation of separate lepton number conservation.
Typically $K_L\to \mu e$ and $K\to \pi\mu e$ are present in models
with tree level flavour-changing neutral current transitions and
are mediated by heavy lepto-quarks, horizontal gauge bosons,
flavour violating neutral scalars etc. The present $90\%$ C.L.
upper bounds are: $Br(K_L\to\mu e)<3.3 \cdot 10^{-11}$ (BNL791),
$Br(K^+\to\pi^+\mu e)<2.1 \cdot 10^{-10}$ (BNL777) and
$Br(K_L\to\pi^0\mu e)<3.2 \cdot 10^{-9}$ (FNAL799). They should
be improved in the next years to: $10^{-12}$ (BNL871),
$2\cdot 10^{-12}$(BNL865) and $10^{-11}-10^{-12}$(KTEV,KAMI)
respectively. These decays probe mass scales as high as a few
100 TeV and consequently can give first information whether there
is some new physics in the "grand desert" between the scales probed
by the colliders of the next decade like LHC and the GUTS scales.

\section{CP-Violation in B-Decays}
CP violation in B decays is certainly one of the most important 
targets of B factories and of dedicated B experiments at hadron 
facilities. It is well known that CP violating effects are expected
to occur in a large number of channels at a level attainable at 
forthcoming experiments. Moreover there exist channels which
offer the determination of CKM phases essentially without any hadronic
uncertainties. Since there exist extensive reviews on this subject
\cite{NQ,B95,RFD}, let me concentrate only on the most important points
and in particular on the recent developments.

\subsection{Strategies for $(\alpha,\beta, \gamma)$}
\subsubsection{$B^0$-Decays to CP Eigenstates}
A time dependent asymmetry in the decay $B^0\to f$ with $f$ being
a CP eigenstate is given by

\begin{eqnarray}
\lefteqn{a_{CP}(t,f) = {\cal A}^{dir}_{CP}(B\to f)\cos(\Delta M t)}
\nonumber \\
&& +
{\cal A}^{mix-ind}_{CP}(B\to f)\sin(\Delta M t)
\label{e8}
\end{eqnarray}
where we have separated the {\it direct} CP-violating contributions 
from those describing {\it mixing-induced} CP violation:
\begin{eqnarray}
{\cal A}^{dir}_{CP}(B\to f) &\equiv& \frac{1-\left\vert\xi_f\right\vert^2}
{1+\left\vert\xi_f\right\vert^2},
\nonumber \\
{\cal A}^{mix-ind}_{CP}(B\to f) &\equiv& \frac{2\mbox{Im}\xi_f}{1+
\left\vert\xi_f\right\vert^2}.
\label{e9}
\end{eqnarray}
In (\ref{e8}), $\Delta M$ denotes the mass splitting of the 
physical $B^0$--$\bar B^0$--mixing eigenstates. 
The quantity $\xi_f$ containing essentially all the information
needed to evaluate the asymmetries (\ref{e9}) 
is given by
\begin{equation}\label{e11}
\xi_f=\exp(-i\phi_M)\frac{A(\bar B\to f)}{A(B \to f)}
\end{equation}
with $\phi_M$ denoting the weak phase in the $B-\bar B$ mixing
and $A(B \to f)$ the decay amplitude. 

Generally several decay mechanisms with different weak and
strong phases can contribute to $A(B \to f)$. These are
tree diagram (current-current) contributions, QCD penguin
contributions and electroweak penguin contributions. If they
contribute with similar strength to a given decay amplitude
the resulting CP asymmetries suffer from hadronic uncertainies
related to matrix elements of the relevant operators $Q_i$.

An interesting case arises when a single mechanism dominates the 
decay amplitude or the contributing mechanisms have the same weak 
phases. Then

\begin{equation}\label{e111}
\xi_f=\exp(-i\phi_M) \exp(i 2 \phi_D),
\quad
\mid \xi_f \mid^2=1
\end{equation}
where $\phi_D$ is the weak phase in the decay amplitude.
In this particular case the hadronic matrix elements drop out,
 the direct CP violating contribution
vanishes and the mixing-induced CP asymmetry is given entirely
in terms of the weak phases $\phi_M$ and $\phi_D$. One has then
\begin{equation}
a_{CP}(f)=\pm  \sin((\Delta M) t)\sin(2\phi_D-\phi_M)
\end{equation}
where $\pm$ refers to $f$ being a $CP=\pm$ eigenstate.
Then one finds for $B_d\to \psi K_S$ and $B_d\to \pi^+\pi^-$
\begin{eqnarray}
 a_{CP}(\psi K_S,t)=-\sin((\Delta M)_d t)\sin(2\beta) 
&&
\nonumber \\
   a_{CP}(\pi^+\pi^-,t)=-\sin((\Delta M)_d t)\sin(2\alpha) 
&&
\label{113c}
\end{eqnarray}
where we have neglected for a moment QCD penguins in $a_{CP}(\pi^+\pi^-,t)$.
Since in the usual unitarity triangle  one side is known,
it suffices to measure
two angles to determine the triangle completely. This means that
the measurements of $\sin 2\alpha$ and $\sin 2\beta$ can determine
the parameters $\varrho$ and $\eta$.

Unfortunately life is not so easy and there are only a few
channels for which this fortunate situation takes place.
In addition studies of this type require tagging ( distinction 
between unmixed $B^0$ and $\bar B^0$ at $t=0$ ) as well as
two time dependent rates $B^0(t)\to f$ and  $\bar B^0(t)\to f$.
Both tagging and time dependent studies are certainly not easy.

\subsubsection{Decays to CP-Non-Eigenstates}
One can of course study also decays to CP non-eigenstates. This,
in addition to tagging, requires generally four time dependent rates 
$B^0(t)\to f$,  $\bar B^0(t)\to f$, $B^0(t)\to \bar f$ and  
$\bar B^0(t)\to \bar f$ \cite{DUN}. In certain cases this approach gives
interesting results. 

\subsubsection{Triangle Constructions}
Here one attempts to extract $\alpha,~\beta,$ and $\gamma$ from
branching ratios only. Neither tagging nor time-dependent
measurements are needed. On the other hand these methods require
measurements of several branching ratios in order to eliminate
the hadronic uncertainties. The prototypes of such studies
are the method of Gronau and Wyler \cite{Wyler} and the
method of Gronau and London \cite{CPASYM} with the latter using the
$SU(2)$-flavour (isospin) symmetry. More recently methods
based on SU(3)-triangle relations have been proposed 
\cite{grl}. I will return to them below.

\subsection{Theoretically Clean Determinations of 
$(\alpha,\beta,\gamma)$}
In my opinion there exist only {\it five} determinations
of $(\alpha,\beta,\gamma)$ in B decays which fully deserve the
name "theoretically clean". Let me discuss them briefly one-by-one.
\subsubsection{$B_d\to\psi K_S$ and $\beta$}
The mixing induced CP-asymmetry in this "gold-plated" decay allows
 in the standard model a direct measurement of the angle $\beta$ as
 pointed out by Bigi and Sanda \cite {BSANDA} a long time ago. 
In this decay the QCD penguins and EW penguins have to an
excellent approximation the same weak
phases as the leading tree contributions. This results in the
formula given in (\ref{113c}) which offers a clean determination
of $\beta$.

\subsubsection{$B_d\to\pi^+\pi^-$ and $\alpha$}
In this case the formula (\ref{113c}) does not really apply
because of the presence of QCD penguin contributions which have
different phases than the leading tree contributions. The
asymmetry $a_{CP}(\pi^+\pi^-)$ measures then $2 \alpha+\theta_P$
where the unknown phase $\theta_P$ signals the presence of
QCD penguins. Using the isospin symmetry and the related triangle
construction Gronau and London \cite{CPASYM} 
have demonstrated how the unknown
phase $\theta_P$ can be found by measuring in addition
the branching ratios $Br(B^+\to \pi^+\pi^0)$, 
$Br(B^0\to \pi^0\pi^0)$ and the branching ratios of CP conjugate
channels. With this information the asymmetry $a_{CP}(\pi^+\pi^-)$
offers a clean determination of $\alpha$. 
The smallness of $Br(B_d\to \pi^0\pi^0$) ($\le{\cal O}(10^{-6})$
\cite{KP}
is a weak point of this method.
It has been pointed out by Deshpande and He \cite{dhewp2} 
that the presence of
EW penguins could have a sizable impact on the GL method.
A closer look shows, however that this impact is rather small 
\cite{ghlrewp},
at most a few $\%$. Moreover a
method has been proposed to estimate this effect quantitatively
\cite{PAPI,PAPIII}.

\subsubsection{$B^{\pm}\to D^0_{CP} K^{\pm}$ and $\gamma$}
This decay involves only tree diagram contributions and requires
six decay rates $B^{\pm}\to D^0_{CP} K^{\pm}$,
$B^+ \to D^0 K^+,~ \bar D^0 K^+$ and  $B^- \to D^0 K^-,~ \bar D^0 K^-$.
A known triangle construction due to Gronau and Wyler \cite{Wyler}
allows then
a clean determination of $\gamma$. The virtue of this method
is that neither tagging nor time-dependent studies are required.
Moreover the observation of CP violation in $B^{\pm}\to D^0_{CP} K^{\pm}$
would signal automatically direct CP violation. A possible difficulty is
the disparity in the size of the expected branching ratios needed 
to construct the triangles in question. Whereas four branching ratios
listed above are expected to be ${\cal O}(10^{-4}-10^{-5})$, the
branching ratio of the colour supressed channels $B^{\pm}\to D^0 K^{\pm}$
are expected to be by one order of magnitude smaller. On one hand
such a small branching ratio is difficult to measure. On the other
hand the resulting triangles will have one side very small making
the extraction of $\gamma$ difficult. A similar method using
neutral B-decays has been proposed by Dunietz \cite{DUN2}.
\subsubsection{$B_s\to D^+_s K^{-}$, $\bar B_s\to D^-_s K^{+}$ and $\gamma$}
This method suggested by Aleksan, Dunietz and Kayser \cite{adk} 
is unaffected
by penguin contributions. A full time dependent analysis allows a clean
measurement of $\sin^2\gamma$. Since the expected branching ratio is
${\cal O}(10^{-4})$, this method is in principle feasible although the 
expected large $B^0_s-\bar B^0_s$ mixing makes this measurement very
challenging.
\subsubsection{$B_s\to\psi\phi$ and $\eta$}
This is an analog of $B_d \to \psi K_s$. In the leading order of the
Wolfenstein parametrization the asymmetry $a_{CP}(\psi\phi,t)$ vanishes.
Including higher order terms in $\lambda$ one finds \cite{B95,DUNS}
\begin{equation}\label{DU}
a_{CP}(\psi\phi,t)=2\lambda^2\eta \sin((\Delta M)_s t)
\end{equation}
where $\lambda$ and $\eta$ are the Wolfenstein parameters.
With $\lambda=0.22$ and $\eta=0.35$ one
has $2\lambda^2\eta=0.03$. The rapid oscillations due to large $(\Delta M)_s$
and the smallness of the CKM factor make the measurement of $\eta$ 
this way  very challenging.
On the other hand the clean character of this asymmetry and its smallness 
can be used in the
search for the physics beyond the standard model.

\subsection{ Four Topics}
\subsubsection{Electroweak Penguins and B decays}
During the last two years there has been a considerable interest in
the role of electroweak penguin contributions in non-leptonic $B$-decays.
Since the Wilson coefficients of the corresponding local operators
increase strongly with the top-quark mass, it has been found by
 Fleischer \cite{rfewp1,rfewp2,rfewp3} that the role of the electroweak
penguins can be substantial in certain decays. This is for instance
the case of the decay $B^-\to K^-\Phi$ \cite{rfewp1},
which  exhibits sizable electroweak penguin effects. 
More interestingly, there are even some channels, such as $B^-\to\pi^-\Phi$
\cite{rfewp2} and $B_s\to\pi^0\Phi$ \cite{rfewp3}, which are 
{\it dominated completely} by electroweak penguin contributions
and which should, thus, allow interesting insights into the physics
of the corresponding operators. In this respect, the
decay $B_s\to\pi^0\Phi$ (or similar transitions such as $B_s\to\rho^0\Phi$)
is very promising due to its special isospin-, CKM- and 
colour-structure \cite{rfewp3}. As the branching ratio of this mode
is expected to be of ${\cal O}(10^{-7})$, it will unfortunately be
rather difficult to analyze this decay experimentally.  
The electroweak penguin effects discussed in \cite{rfewp1,rfewp3}
have been confirmed by other authors \cite{dhewp}-\cite{dy}.

Now the question arises whether
the usual strategies for the determination of the CKM-phases are
affected by the presence of the electroweak penguin contributions.

The five clean strategies reviewed above are,
 -- except for the Gronau-London method -- 
unaffected by EW penguins. It can
also be shown that the impact of the $\bar b \to \bar d$ EW penguins 
on the Gronau-London method is small \cite{ghlrewp,PAPI}. 
On the other hand as pointed out by Despande and He \cite{dhewp2},
the $\bar b \to \bar s$ EW penguins have a considerable impact on 
the GHLR method \cite{grl}. This method uses 
the $SU(3)$ flavour symmetry 
of strong interactions in
B-decay into $\pi\pi$, $\pi K$ and $K\bar K$ final states which 
should allow to determine the CKM phases  
by measuring
only branching ratios of the relevant B decays.
In view of this problem various strategies have been proposed
in order to either eliminate electroweak penguins
\cite{ghlrewp,dhgam} or to determine them \cite{PAPI}.
Some of these strategies show sensitivity to $SU(3)_F$ breaking
effects and generally involve many channels. Consequently
they are very challenging for experimentalists and their
usefulness is unclear at present.

Finally I would like to mention 
\cite{PAPIII} where various strategies for fixing the angle $\gamma$
and obtaining experimental insight into the world of EW-penguins have been
presented.
 A by-product of this work is a refined estimate of the role
of EW-penguins in the $\alpha$-determination by means of $B\to\pi\pi$
decays.
 \subsubsection{The Quest for $\alpha$}
The Gronau-London method for the determination of $\alpha$ involves the
experimentally difficult mode $B_d \to \pi^0\pi^0$ which is
expected to be $\le {\cal O}(10^{-6})$ \cite{KP}.
For this reason several other methods have been proposed which
avoid the use of $B_d \to \pi^0\pi^0$ in conjunction with
$a_{CP}(\pi^+\pi^-,t)$. For instance in \cite{SNYD} the use of 
$B \to \varrho \pi$ has been suggested.
Next in \cite{PAPII} an alternative method of
determining $\alpha$ by performing simultaneous measurements
of the mixing-induced CP asymmetries of the decays 
$B_d\to\pi^+\pi^-$ and $B_d\to K^0\bar K^0$ has been proposed.
 The accuracy of
this method is limited by $SU(3)$ breaking effects but it is
unaffected by EW penguins. Moreover the decay $B_d \to K^0\bar K^0$
might be easier to measure than $B_d \to \pi^0\pi^0$. 
As shown by Fleischer \cite{PAPF} the CP asymmetry in
$B_d \to K^0\bar K^0$ can be as large as ${\cal O}(50\%)$.
Next in \cite{Gronau}
the use of $a_{CP}(\pi^+\pi^-,t)$ together with the
rates for B-meson decays into $\pi^{\pm}K^{\mp}$ and $K_S\pi^{\pm}$
allows with the help of $SU(3)_F$ symmetry (including its first-order
symmetry breaking) the extraction of $\alpha$ and $\gamma$.
Finally as pointed out in \cite{FMANNEL}, $\alpha$ can be determined
by measuring $A_{CP}^{dir}(B^0\to\pi^+\pi^-)$,
$A_{CP}^{mix-ind}(B^0\to\pi^+\pi^-)$, $R_t$ and 
$Br(B^+\to\pi^+K^0)/Br(B^+\to\pi^+\pi^0)$.

On the other hand, to sort out the effects of QCD penguins in
$B^0\to \pi^+\pi^-$, 
Silva and Wolfenstein \cite{SW94} suggest to measure
$\Gamma(\bar B^0\to\pi^+K^-)/\Gamma(\bar B^0\to\pi^+\pi^-)$
with the first channel dominated by penguin contributions.
Next DeJongh and Sphicas \cite{DJS}, 
Kramer, Palmer and Wu \cite{KPW}, and 
Aleksan et al. \cite{ALEKSAN} 
provide various model estimates of the importance
of QCD penguins in $a_{CP}(\pi^+\pi^-,t)$. In particular Kramer
et al. expect only a small impact of QCD penguins on the determination
of $\alpha$ from $a_{CP}(\pi^+\pi^-,t)$ provided $\alpha$ is large.
Different conclusion has been reached by Gronau \cite{GRON}. Certainly
the chapter on $\alpha$ determination from $a_{CP}(\pi^+\pi^-,t)$
is not closed.
\subsubsection{Time-dependent Untagged Strategies}
The expected large $B^0_s-\bar B^0_s$ mixing implies rapid
oscillations in $a_{CP}(f,t)$ which complicates the extraction
of weak phases from $B_s$ decays by means of tagged measurements.
As pointed out by Dunietz \cite{D1} and analyzed in detail by him and 
Fleischer \cite{FD12}, 
these rapid oscillations cancel in untagged measurements
which in addition are more efficient experimentally than the tagged
ones. The size of CP violating effects is governed then by the width
difference $(\Delta\Gamma)_s$ in the $B^0_s-\bar B^0_s$ system,
which is known to be substantial. Using various angular distributions
of decay products, clean determinations of $\gamma$ in
$B_s\to D^{*\pm}K^{*\mp},~K^{*+}K^{*-},~K^{~0}\bar K^{~0}$ and of $\eta$
in $B_s\to \psi\phi$ are then possible \cite{FD12}.

The feasibility of this method depends on the actual size of
$(\Delta\Gamma/\Gamma)_{B_s}$ and the method ceases to be useful if
this ratio is well below 0.2. Unfortunately a recent calculation 
\cite{BBD1} shows 
that the inclusion of ${\cal O}(\Lambda_{QCD}/m_b)$ and
${\cal O}(\ms/\mb)$ corrections reduces this ratio by roughly $30\%$
relative to previous estimates:
$(\Delta\Gamma/\Gamma)_{B_s}=0.16+0.11-0.09$. The inclusion of
short distance NLO corrections and better estimates of the relevant
non-perturbative $B_i$ factors could reduce the sizable error.
Eventually $(\Delta\Gamma)_{B_s}$ should be measured at HERA-B and in
Run II at FNAL.
\subsubsection{CP Asymmetries in Inclusive Decays}
The CP violating asymmetries that occur in partially inclusive
neutral B meson decays with final states specified by their flavour
content have been studied recently \cite{BBD2}. Such inclusive
asymmetries are complementary to those in exclusive decays and can
be sometimes theoretically advantageous. In particular the
asymmetry in $B_d$ meson decay into charmless final states with
no net strangeness provides a determination of $\sin 2\alpha$,
that does not suffer from uncontrolled penguin contributions
or tiny branching ratios. Finally recent studies of CP asymmetries
in $B_d\to X_{s,d}\gamma$ and $B\to X_d l^+l^-$ can be found
in \cite{Asatrian} and \cite{Sehgal96} respectively. 
CP-violating asymmetries in $B\to K^*(\varrho)\gamma$ are 
presented in \cite{GRWS}.
\section{Future Visions}
\subsection{Classification}
Let us begin this section by grouping various 
decays and 
quantities into four distinct classes with respect to theoretical
uncertainties. 
\subsubsection{Gold-Plated Class}
These are the decays with essentially no theoretical uncertainties:
\begin{itemize}
\item
CP asymmetries in $B_d\to \psi K_S$ and  $B_s\to\psi\phi$ which
measure the angle $\beta$ and the parameter $\eta$ respectively,
\item
The ratio  $Br(B\to X_d\nu\bar\nu)/ Br(B\to X_s\nu\bar\nu)$
which offers the cleanest direct determination of the ratio 
$|V_{td}/V_{ts}|$,
\item
Rare K-decays $K_L \to \pi^0\nu\bar\nu$ and $K^+\to \pi^+\nu\bar\nu$
which offer very clean determinations of $\IM\lambda_t (\eta)$ and
$|V_{td}|$ respectively.
\end{itemize}
\subsubsection{Class 1}
\begin{itemize}
\item
CP asymmetry in $B^0\to \pi^+\pi^-$ relevant for the angle $\alpha$ and the
CP asymmetries in $B^{\pm}\to D_{CP}K^{\pm}$,
$B_s\to D_sK$ and $B^0\to \bar D^0K^*$ all relevant for the angle $\gamma$.
These CP asymmetries
require additional strategies in order to determine these angles without
hadronic uncertainties.
\item
Ratios $Br(B_d\to l\bar l)/Br(B_s\to l\bar l)$ and
$(\Delta M)_d/(\Delta M)_s$
which give  good measurements of $|V_{td}/V_{ts}|$ 
provided the SU(3) breaking effects in the ratios $F_{B_d}/F_{B_s}$
and $\sqrt{B_d}F_{B_d}/\sqrt{B_s}F_{B_s}$ can be brought under control.
\end{itemize}
\subsubsection{Class 2}
Here I group  quantities or decays with presently moderete ($\pm 10\%$)
 or substantial ($\pm 20\%$)
theoretical uncertainties which should be considerably reduced in the next
five years. In particular I assume that the uncertainties in $B_K$
and $\sqrt{B}F_B$ will be reduced below 10\%.
\begin{itemize}
\item
$B\to X_{s,d}\gamma$, $B\to X_{s,d} l^+ l^-$, $B\to K^*(\rho)l^+l^-$
\item
$(\Delta M)_d$, $(\Delta M)_s$, $\vcb_{excl}$, 
$\vcb_{incl}$,
$| V_{ub}/V_{cb}|_{incl}$
\item
Some CP asymmetries in B-decays discussed above
\item
$\varepsilon_K$ and $K_L\to \pi^0 e^+e^-$
\end{itemize}
\subsubsection{Class 3}
Here we have a list of important decays with large theoretical
uncertainties which can only be removed by a dramatic progress
in non-perturbative techniques:
\begin{itemize}
\item
CP asymmetries in most $B^{\pm}$-decays
\item
$B_d\to K^*\gamma$, Non-leptonic B-decays, $| V_{ub}/V_{cb}|_{excl}$
\item
$\varepsilon^{\prime} /\varepsilon$, $K\to \pi\pi$, $\Delta M(K_L-K_s)$,
$K_L\to\mu\bar\mu$, hyperon decays and so on.
\end{itemize}
It should be stressed that even in the presence of theoretical
uncertainties a measurement of a non-vanishing 
ratio $\varepsilon^{\prime}/\varepsilon$ or a non-vanishing CP asymmetry
in charged B-decays would signal direct CP violation excluding
superweak scenarios \cite{WO1}. This is not guaranteed by
several clean decays of the gold-plated class or class 1 \cite{WIN} except for 
$B^{\pm}\to D_{CP} K^{\pm}$.
\subsection{Some Numerical Examples}
In what follows let us assume that the problems with the determination
of $\alpha$ discussed in the previous section 
will be solved somehow. Since in the usual unitarity triangle
 one side is known, it suffices to measure
two angles to determine the triangle completely. This means that
the measurements of $\sin 2\alpha$ and $\sin 2\beta$ can determine
the parameters $\varrho$ and $\eta$.
As the standard analysis of the unitarity triangle of section 3
shows, $\sin(2\beta)$ is expected to be large: $\sin(2\beta)=0.58\pm 0.22$
implying the integrated asymmetry  $A_{CP}(\psi K_S)$
as high as $(30 \pm 10)\%$.
The prediction for $\sin(2\alpha)$ is very
uncertain on the other hand $(0.1\pm0.9)$ and even a rough measurement
of $\alpha$ would have a considerable impact on our knowledge of
the unitarity triangle as stressed in \cite{BLO} and recently in
\cite{BB96}.

Let us then compare the potentials of the CP asymmetries in
determining the parameters of the standard model with those
of the cleanest rare K-decays: $K_L\to\pi^0\nu\bar\nu$ and
$K^+\to\pi^+\nu\bar\nu$ \cite{BB96}.
Measuring $\sin 2\alpha$ and $\sin 2\beta$ from CP asymmetries in
$B$ decays allows to fix the 
parameters $\bar\eta$ and $\bar\varrho$ \cite{BLO,B94}.
Alternatively, $\bar\varrho$ and $\bar\eta$ may also be determined
from $K^+\to\pi^+\nu\bar\nu$ and $K_L\to\pi^0\nu\bar\nu$ alone
\cite{BH,BB4}. An interesting feature of this possibility is the
fact that the extraction of $\sin 2\beta$ from these
two modes is essentially independent of $\mt$ and $V_{cb}$
\cite{BB4}. This enables a rather accurate determination of
$\sin 2\beta$ from $K\to\pi\nu\bar\nu$.

A comparison of both strategies
is displayed in tables 4 and 5, where 
the following input has been used:
$|V_{cb}|=0.040\pm 0.002(0.001)$, $\mt=(170\pm 3) \gev$
and
\begin{eqnarray}
B(K_L\to\pi^0\nu\bar\nu) = (3.0\pm 0.3)\cdot 10^{-11}
&&
\nonumber \\
B(K^+\to\pi^+\nu\bar\nu) = (1.0\pm 0.1)\cdot 10^{-10}
&&
\label{bklkp}
\end{eqnarray}
\\
The measurements of CP asymmetries in $B_d\to\pi\pi$ and
$B_d\to J/\psi K_S$, expressed in terms of $\sin 2\alpha$ and
$\sin 2\beta$, are taken to be \\
scenario I:
\begin{equation}\label{sin2a2bI}
\sin 2\alpha=0.40\pm 0.10 \quad \sin 2\beta=0.70\pm 0.06
\end{equation}
scenario II:
\begin{equation}\label{sin2a2bII}
\sin 2\alpha=0.40\pm 0.04 \quad \sin 2\beta=0.70\pm 0.02
\end{equation}
Scenario I corresponds to the accuracy being aimed for at $B$-factories
and HERA-B prior to the LHC era. An improved precision can be anticipated from
LHC experiments, which we illustrate with our choice of scenario II.

As can be seen in tables 4 and 5, the CKM determination
using $K\to\pi\nu\bar\nu$ is competitive with the one based
on CP violation in $B$ decays, except for $\bar\varrho$ which
is less constrained by the rare kaon processes.

\begin{table}
\begin{center}
\begin{tabular}{|c||c|c|}\hline
&$K\to\pi\nu\bar\nu$&CP-B (I)\\ 
\hline
\hline
$|V_{td}|/10^{-3}$&$10.3\pm 1.1$&$8.8\pm 0.5$\\ 
\hline
$|V_{ub}/V_{cb}|$&$0.089\pm 0.017$
&$0.087\pm 0.009$ \\
\hline 
$\bar\varrho$&$-0.10\pm 0.16$ &$0.07\pm 0.03$\\
\hline
$\bar\eta$&$0.38\pm 0.04$&$0.38\pm 0.04$\\
\hline
$\sin 2\beta$&$0.62\pm 0.05$&$0.70\pm 0.06$\\
\hline
${\rm Im}\lambda_t/10^{-4}$&$1.37\pm 0.07$
&$1.37\pm 0.19$ \\
\hline
\end{tabular}
\end{center}
\caption[]{Illustrative example of the determination of CKM
parameters from $K\to\pi\nu\bar\nu$ and from CP violating
asymmetries in $B$ decays (scenario I).
$V_{cb}=0.040\pm 0.002$.
\label{tabkb1}}
\end{table}

\begin{table}
\begin{center}
\begin{tabular}{|c||c|c|}\hline
&$K\to\pi\nu\bar\nu$&CP-B (II)\\ 
\hline
\hline
$|V_{td}|/10^{-3}$&$10.3\pm 0.9$
&$8.8\pm 0.2$ \\
\hline
$|V_{ub}/V_{cb}|$&$0.089\pm 0.011$
&$0.087\pm 0.003$ \\
\hline 
$\bar\varrho$&$-0.10\pm 0.12$
&$0.07\pm 0.01$ \\
\hline
$\bar\eta$&$0.38\pm 0.03$
&$0.38\pm 0.01$ \\
\hline
$\sin 2\beta$&$0.62\pm 0.05$
&$0.70\pm 0.02$ \\
\hline
${\rm Im}\lambda_t/10^{-4}$&$1.37\pm 0.07$
&$1.37\pm 0.08$ \\
\hline
\end{tabular}
\end{center}
\caption[]{Illustrative example of the determination of CKM
parameters from $K\to\pi\nu\bar\nu$ and from CP violating
asymmetries in $B$ decays ( scenario II). 
$V_{cb}=0.040\pm 0.001$.
\label{tabkb2}}
\end{table}

On the other hand ${\rm Im}\lambda_t$ is better determined
in the kaon scenario. It can be obtained from
$K_L\to\pi^0\nu\bar\nu$ alone and does not require knowledge
of $V_{cb}$ which enters ${\rm Im}\lambda_t$ when derived
from $\sin 2\alpha$ and $\sin 2\beta$.
This analysis suggests that $K_L\to\pi^0\nu\bar\nu$ should eventually 
yield the most accurate value of ${\rm Im}\lambda_t$.
This would be an important result since ${\rm Im}\lambda_t$
plays a central role in the phenomenology of CP violation
in $K$ decays and is furthermore equivalent to the 
Jarlskog parameter $J_{CP}$ \cite{CJ}, 
the invariant measure of CP violation in the Standard Model, 
$J_{CP}=\lambda(1-\lambda^2/2){\rm Im}\lambda_t$.

There is another virtue of the comparision of the determinations
of various parameters using CP-B asymmetries with the determinations
in very clean decays $K\to\pi\nu\bar\nu$. Any substantial deviations
from these two determinations would signal new physics beyond the
standard model.

On the other hand  
unprecedented precision for all CKM
parameters could be achieved by combining the cleanest K and 
B decays \cite{B94} . 
While $\lambda$ is obtained as usual from
$K\to\pi e\nu$, $\bar\varrho$ and $\bar\eta$ could be determined
from $\sin 2\alpha$ and $\sin 2\beta$ as measured in CP
violating asymmetries in $B$ decays. Given $\eta$, one could
take advantage of the very clean nature of $K_L\to\pi^0\nu\bar\nu$
to extract $A$ or, equivalently $|V_{cb}|$ using (\ref{bklpn}). 
For $\sin 2\alpha=0.40\pm 0.04$,
$\sin 2\beta=0.70\pm 0.02$ and 
$B(K_L\to\pi^0\nu\bar\nu)=(3.0\pm 0.3)\cdot 10^{-11}$,
$m_t=(170\pm 3)GeV$ one finds 
\begin{eqnarray}
&&
\bar\varrho=0.07\pm 0.01 \qquad
\bar\eta=0.38\pm 0.01
\nonumber \\
&&
|V_{cb}|=0.0400\pm 0.0013
\label{rhetvcb}
\end{eqnarray}
which would be a truly remarkable result. Again the comparision of
this determination of $|V_{cb}|$ with the usual one in tree level
B-decays would offer an excellent test of the standard model
and in the case of discrepancy would signal physics beyond the
standard model.  

\section{A Look beyond the Standard Model}
In this  review we have concentrated on rare decays and  CP violation in the
standard model. The structure of rare decays and of CP violation in 
extensions of the 
standard model may deviate from this picture.
Consequently the situation in this field could turn out to be very different
from the one presented here. It is appropriate then to end this review with
a few remarks on the physics beyond the standard model. Much more
elaborate discussion can be found for instance in \cite{NIRNEW,Ligetti},
where further references can be found.
\subsection{Impact of New Physics}
There is essentially no impact on $|V_{us}|$, $|V_{cb}|$ and $|V_{ub}|$
determined in tree level decays. This is certainly the case for the first
two elements. In view of the smallness of $|V_{ub}|$ a small impact from
the loop contributions (sensitive to new physics) to leading decays
could in principle be present. However in view of many theoretical
uncertainties in the determination of this element such contributions
can be safely neglected at present.

There is in principle a substantial impact of new physics on the
determination of $\varrho$, $\eta$, $|V_{td}|$, $\IM\lambda_t$ and
generally on the unitarity triangle through the loop induced
decays which can receive new contributions from internal chargino, 
charged Higgs, stops, gluinos and other exotic exchanges. If the
quark mixing matrix has the CKM structure, the element $|V_{ts}|$
on the other hand will be only slightly affected by these new
contributions. Indeed from the unitarity of the CKM matrix
$|V_{ts}|/|V_{cb}|=1-{\cal O}(\lambda^2)$ and the new contributions
could only affect the size of the ${\cal O}(\lambda^2)$ terms which
amounts to a few percent at most. This situation makes the study
of new physics in rare B decays governed by $|V_{ts}|$ somewhat
easier than in rare K-decays and B decays which are governed by
$|V_{td}|$. Indeed in the latter decays the impact of new physics
is felt both in the CKM couplings and in the $\mt$ dependent functions,
which one has to disantangle, whereas in the former decays
mainly the impact of new physics on the $m_t$ dependent functions
is felt.

Similarly if no new phases in the quark mixing are present, 
the formulae for CP asymmetries in B-decays remain unchanged and these
asymmetries measure again the phases of the CKM matrix as in the
standard model. Thus even if there is some new physics in the loop
diagrams we will not see it in the clean asymmetries directly if
there are no new phases in the quark mixing matrix. In order
to search for new physics  and  to be able to distinguish between
various types of new physics, the comparision of the values of CKM phases
determined from CP asymmetries and 
from loop induced decays is then mandatory.

The situation becomes even more complicated if the quark mixing involves
more angles and new phases and in addition there are new parameters 
in the Higgs, SUSY and generally new physics sector. For instance
in such a case the "gold-plated" asymmetry in $B \to\psi K_S$ 
would take the form \cite{NIRNEW}:
\begin{equation}\label{113u}
 A_{CP}(\psi K_S,t)=-\sin((\Delta M)_d t)\sin(2\beta+\theta_{NEW}) 
\end{equation}
implying that not $2\beta$ but  $2\beta+\theta_{NEW}$ is measured by
the asymmetry. Some strategies for the extraction of phases in
these more complicated situations have been recently discussed
\cite{GL96,W96} 
 
This short discussion makes it clear that in order to search effectively
for new physics it is essential to measure and calculate as many
processes and compare the resulting CKM parameters with each other.
Graphically this corresponds simply to figure 3. In this enterprise
the crucial role will be played by very clean decays of the "gold-plated"
class and of classes 1 and possibly 2 in which the new physics will
not be hidden by theoretical uncertainties present in the decays of class 3.

\subsection{Signals of New Physics}
New Physics will be signalled in principle in various ways. 
Here are some obvious examples:
\begin{itemize}
\item
Standard model predictions for various branching ratios and CP
asymmetries will disagree with data,
\item
$(\varrho,\eta)$  determined in K-physics will disagree with
$(\varrho,\eta)$ determined in B-physics,
\item
$(\varrho,\eta)$ determined in loop induced decays will disagree
with $(\varrho,\eta)$ determined through CP asymmetries,
\item
Forbidden and very rare decays will occur at unexpected level:
$K_L\to \mu e$, $K\to \pi \mu e$, $d_N$, $d_e$, $\mu\to e\gamma$,
$D^0-\bar D^0$
mixing, CP violation in D-decays \cite{Burdman} etc.,
\item
Unitarity Triangle will not close.
\end{itemize}
\subsection{General Messages}
Let us end this discussion with some general messages on New Physics.

It is well known that baryogenesis suggests some CP violation
outside the Standard Model. The single CKM phase simply does not give
enough CP Violation for the required baryon asymmetry \cite{Gavela}. 
It is however not unlikely that large new sources of
CP violation necessary for baryogenesis could be present at the electroweak 
scale \cite{Nelson}. They 
are present for instance in general SUSY models and in multi-Higgs models.

It should be stressed that baryogenesis and the required additional CP
violation being flavour diagonal may have direct impact on the electric
dipole moments but have no direct impact on FCNC processes. However
new physics required for baryon asymmetry could bring new phases
relevant for FCNC.

Concentrating on SUSY for a moment, more general and natural 
SUSY models give
typically very large CP violating effects \cite{Gerar} and FCNC transitions
 \cite{Nilles} which
are inconsistent with the experimental values of $\varepsilon_K$, 
$K_L-K_S$ mass difference and the bound on the electric dipole moment
of the neutron. In order to avoid such problems, special forms of 
squark mass matrices \cite{Masiero}
and fine tunning of phases are necessary. In addition
one frequently assumes that CP violation and FCNC are absent at tree
level. 
In the limiting case one ends with a special version of 
the MSSM in which to a good
approximation CP violation and FCNC processes are governed by the
CKM matrix and the new effects are dominantly described by loop diagrams
with internal stop, charginos and charged higgs exchanges.
It is then not surprising that in the quark sector new effects in MSSM 
compared with SM predictions for FCNC transitions 
are rather moderate, although for a particular
choice of parameters and certain quantities still enhancements 
(or suppressions)
by factors 2-3 cannot be excluded \cite{Branco,ASYMM}.
This turns out to be the case e.g. for $\varepsilon_K$ and
$(\Delta M)_{d,s}$. Unfortunately in view of the uncertainties
related to the non-perturbative factors like $B_K$ and $\sqrt{B}F_B$
such "moderate" effects will not be easily seen. 
Larger effects are expected in the lepton
sector and in electric dipole moments \cite{BARBIERI}. 
Similar comments about the size
of new effects apply to multi-higgs models and left-right symmetric
models.

Large effects are still possible in models with tree level FCNC
transitions \cite{Reina,Ligetti}, 
leptoquarks, models with horizontal gauge symmetries,
technicolour and top-colour models \cite{NIRNEW,Ligetti}. 
Unfortunately these 
models
contain many free parameters and at present the main thing  one
can do is to bound numerous new couplings and draw numerous
curves which from my point of view is not very exciting.

On the other hand it is to be expected that  clearest signals
of new physics may come precisely from very exotic physics
which would cause the decays $K_L\to \mu e$, $K\to \pi \mu e$, 
T-violating $\mu$-polarization in $K^+\to \pi^0\mu^+\nu$ to occur. Also
sizable values of $d_N$, $d_e$, $\mu\to e \gamma$, $D^0-\bar D^0$
mixing, $D^0\to l^+l^-$ and of CP violation in D-decays and 
top decays are very interesting
in this respect. Important progress in FCNC transitions and CP
violation in the
D-system should be achieved at FNAL at the beginning of the next 
decade \cite{Kaplan96}.
In the standard models these effects are very small but in a number
of extensions like SUSY models, multiscalar models and leptoquark
models the new contribution could bring CP violation and
$D^0-\bar D^0$ mixing close to present experimental limits
\cite{DDBAR,Burdman}. 

It should however be  stressed once more that theoretically
cleanest decays belonging to the top classes of section 8 will
certainly play important roles in the search for new physics
and possibly will offer its first signals.
\section{Summary and Outlook}
During the recent years important progress has been made in
the field of FCNC processes through:

\begin{itemize}
\item
The improved measurement of $\mt$.
\item
The improved determinations of $\vcb$ and $|V_{ub}|$.
\item
The discovery of $b\to s \gamma$ transition and the improved
bounds on several branching ratios for rare decays.
\item
The calculation of NLO short distance QCD corrections for most
important decays.
\item
Heavy Quark Expansions and Heavy Quark Effective Theory.
\item
Improved non-perturbative calculations.
\item
Suggesting of useful methods for extracting the parameters of
the standard model with small or negligible theoretical uncertainties.
\end{itemize}

The standard model is fully consistent with the existing data for
FCNC processes. The next ten years could change this picture
dramatically through the advances in experiment and theory.
In particular:

\begin{itemize}
\item
The error on $\vcb$ and $\vub$ could be decreased below 0.002
and 0.01 respectively. This progress should come mainly from
Cornell, B-factories and new theoretical efforts.
\item
The error on $\mt$ should be decreased down to $\pm 3\gev$
at Tevatron and $\pm 1\gev$ at LHC.
\item
The improved measurements of $\epe$ should give some insight
in the physics of direct CP violation inspite of large
theoretical uncertainties.
\item
The first events for $K^+\to\pi^+\nu\bar\nu$ could in principle
bee seen at BNL already next year. A detailed study of this very
importrant decay requires however new experimental ideas and
new efforts.
\item
The future improved inclusive $b \to s \gamma$ measurements
confronted with improved standard model predictions could
give the first signals of new physics.
\item
Similarly the measurement of the transition $b \to s \mu\bar\mu$,
to be expected at the beginning of the next decade, should be
very important in this respect.
\item
The theoretical status of $K_L\to \pi^0 e^+ e^-$ and 
$K_L\to \mu\bar\mu$ should be improved to confront future
data.
\item
The measurement of the $B_s-\bar B_s$ mixing and in particular of
$ K_L\to \pi^0\nu\bar\nu$, $B \to X_{s,d}\nu\bar\nu$ and 
$B_{s,d}\to \mu\bar\mu$ will take most probably longer time but
as stressed in this review all efforts should be made to measure
these transitions.
\item
Clearly future B-factories, HERA-B, LHC-B and the precise
measurements of $\alpha$, $\beta$ and $\gamma$ may totally
revolutionize this field. In particular the first signals
of new physics could be found in the $(\bar\varrho,\bar\eta)$ plane.
\item
The forbidden or strongly suppressed transitions such as
$D^0-\bar D^0$ mixing and $K_L\to \mu e$ are also very
important in this respect
\item
One should hope that the non-perturbative
methods will be considerably improved.
\end{itemize}

In any case the FCNC transitions have a great future and
I expect that they could dominate our field in the first part
of the next decade. 

{\bf Acknowledgements}: 
This work has been supported by the
German Bundesministerium f\"ur Bildung and Forschung under contract 
06 TM 743 and DFG Project Li 519/2-1.

\section*{Questions}
\noindent{\it G. Farrar, Rutgers Univ.:}

Are there any theoretical or experimental developments on FCNC
in the D-meson sector following up on theoretical suggestions of
a few years ago that non-SM physics could give effects just below 
existing limits?

\vskip 12pt
\noindent{\it A.J. Buras:}

Yes, they are discussed at the end of section 9.

\vskip 12pt
\noindent{\it L. Wolfenstein, Carnegie Mellon:}

Does the lattice measurement of a low value of $\ms$ really tell
us the value of $\epe$ or should we get a direct determination
of the matrix elements of $Q_6$ and $Q_8$ from the lattice?

\vskip 12pt
\noindent{\it A.J. Buras:}

The  values of $\ms$ from the lattice measurements have
 certainly implications
on $\epe$ provided the relevant $B_i$ factors are calculated
consistently. On the other hand,
I expect that 
future lattice calculations will directly give the relevant hadronic 
matrix elements and the issue of $\ms$ in connection with $\epe$ will
disappear. 

\vfill\eject

\end{document}